\documentclass{aa}

\usepackage{graphicx}
\usepackage{xcolor}
\usepackage{threeparttable}

\usepackage{txfonts}
\usepackage{mathtools}

\usepackage{hyperref}
\hypersetup{
  colorlinks=true,
  citecolor=blue,
  linkcolor=blue!80!black,
  urlcolor=blue!80!black
}

\usepackage[utf8]{inputenc}
\usepackage[T1]{fontenc}
\usepackage{babel}

\usepackage{float}
\usepackage{placeins}

\usepackage{eso-pic}
\AddToShipoutPictureFG{%
  \AtPageLowerLeft{%
    \put(\LenToUnit{\dimexpr\paperwidth-7mm\relax},
         \LenToUnit{0.5\paperheight}){%
      \rotatebox{90}{%
        \makebox[0pt][c]{%
          \color{gray}\large
          ACCEPTED for publication in Astronomy \& Astrophysics on November 20, 2026%
        }%
      }%
    }%
  }%
}

\begin{document}
\raggedbottom

%======================================================================
% Title, authors, abstract, keywords
%======================================================================
\title{Tracking low-velocity ejecta from the DART impact on Dimorphos}

\titlerunning{Tracking the low-velocity ejecta from the DART impact}

\author{
  Isabel Herreros\inst{1,3}
  \and
  Sébastien Charnoz\inst{2}
}

\authorrunning{I. Herreros \& S. Charnoz}

\institute{
  Centro de Astrobiologia CSIC-INTA, Carretera Ajalvir Km. 4,
      28850 Torrejon de Ardoz, Madrid, Spain;
      \email{iherreros@cab.inta-csic.es}
  \and
  Université Paris Cité, Institut de Physique du Globe de Paris, CNRS,
      F-75005 Paris, 1 rue Jussieu, 75005 Paris, France;
      \email{charnoz@ipgp.fr}
  \and
  Departamento de Ingeniería Térmica y Fluidos,
      Universidad Carlos III de Madrid, 28911 Leganés, Madrid, Spain
}

\abstract
  {The DART impact on Dimorphos produced a large population of low-velocity
  ejecta ($v < v_{\rm esc}$), likely containing most of the excavated mass,
  whose early fate remains poorly constrained.}
  {We investigate the first $\sim 22$~h evolution of ejecta launched at
  $1$--$9$~cm\,s$^{-1}$ and assess how orbital dynamics and {surface transport after first contact {shape  the surface distribution of the re-accreted ejecta blanket.}}}
  {We use {RAVEL}, a model that couples three-dimensional orbital
  dynamics in the Didymos--Dimorphos system with surface transport on a
  digital terrain model, including detachment, re-impact, rebound, and frictional sliding.}
  {Re-accretion is rapid and asymmetric: more than 99\% of the re-accreted
  mass returns to Dimorphos within $\sim 5$~h. The slowest ejecta remain
  concentrated near the DART crater and dominate the primary ejecta blanket,
  whereas faster particles undergo orbital transport and preferentially
  populate antipodal and trailing regions. Surface motion strongly modifies
  the first-contact pattern, and the DART-derived rough terrain model
  produces {elongated, topographically channelled depositional streaks dominated by the slowest ejecta{, giving rise to a ray-like pattern.}}}
  {These results provide testable predictions for ESA's \textit{Hera}
  mission and link ejecta-blanket morphology to {orbital dynamics, re-accretion, and surface topography}.}

\maketitle
\nolinenumbers


%======================================================================
\section{Introduction}
\label{sec:introduction}

NASA's \textit{Double Asteroid Redirection Test} (DART) impact on Dimorphos
on 26 September 2022 provided the first full-scale asteroid-deflection
experiment \citep{Daly_2023_Dart_Nature, Cheng_2023}. The impact shortened Dimorphos' orbit by
$\sim 33$~min \citep{Cheng_2023} and generated an ejecta plume spanning a
broad range of particle sizes and velocities
\citep{Hirabayashi_2025,Li_2023}.

{Previous studies have investigated different stages of the evolution of DART ejecta. {\cite{Farnham_fast_boulders_2025} and \cite{Deshapriya_dart_ejecta_2026} focus mainly on the early ejecta plume, whereas
\cite{Langner_2024,Langner_2025}} investigate the longer-term orbital evolution and late re-impacts ($>24$~h) of DART ejecta in the Didymos--Dimorphos system, focusing mainly on relatively fast ejecta ($\gtrsim 6$~cm\,s$^{-1}$) as they contribute
to the momentum variation of Dimorphos, the target of the DART mission.
The latter studies showed that} particles may re-impact Didymos or Dimorphos, remain
temporarily bound, or escape. However, the re-accretion of low-velocity
ejecta (with velocities smaller than or comparable to the escape velocity of
Dimorphos, $\sim 9$~cm\,s$^{-1}$), expected to dominate the re-accreted
mass (Fig.~\ref{fig:mass_ejected_powerlaw}), remains largely unexplored
and is the subject of the present work.

By the time ESA's \textit{Hera} mission reaches the system, most of the
DART ejecta that remains bound to the Didymos--Dimorphos system is expected
to have re-accreted. {Understanding its redistribution over Dimorphos is therefore essential to interpret Hera observations and, more generally, to improve our understanding of surface mass redistribution on low-gravity rubble-pile bodies.}

Here we follow the first $\sim 22$~h ($\sim 2$ orbits) of ejecta launched
at $1$--$9$~cm\,s$^{-1}$ using {RAVEL}
(\textit{Regolith Astrodynamics in Variable Effective Low-gravity
environments}), a code that continuously couples orbital dynamics with
surface transport after return to Dimorphos, using a Dimorphos digital
terrain model (DTM), while simultaneously integrating the orbital motion
and rotation of Dimorphos. The model traces ejecta from launch to final
deposition, including detachment, re-impact, rebound, and frictional
sliding at the asteroid surface.

To understand the effect of global surface roughness on the final
deposition of ejecta at the surface of Dimorphos, we first use a smooth
triaxial ellipsoid DTM as a reference case, then a DART/DRACO-based
topographically rough DTM (in which only the imaged hemisphere is resolved;
\cite{Daly_2024}), and finally a synthetic fully rough DTM
(Sect.~\ref{sec:fully_rough_DTM}). {We show that the coupling between ejecta dynamics and surface topography can organize slow re-accreted ejecta into elongated, topographically channelled depositional structures whose locations and morphology depend on the underlying terrain. { These structures may be visible as ray-like patterns around the crater.}}

{This paper is organized as follows. In Sect.~\ref{sec:method} we describe the numerical procedure used to follow ejecta dynamics, coupling orbital and surface motion in RAVEL. In Sect.~\ref{sec:results_smooth} we present the standard smooth-ellipsoid case, and in Sect.~\ref{sec:DART_DTM} we study the ejecta distribution on the DART/DRACO-based shape model. We then explore the sensitivity to the spatial extent of the ejecta source in Sect.~\ref{sec:radius_15m}, the effect of globally distributed roughness in Sect.~\ref{sec:fully_rough_DTM}, and the sensitivity to the assumed launch position--velocity relation in Sect.~\ref{sec:random_initial_conditions}, before summarizing the main conclusions.}
%======================================================================
\section{Method}
\label{sec:method}

{RAVEL}  is a custom-developed code designed to track the coupled orbital and
surface motion of low-velocity ejecta on small bodies. The model tracks
Lagrangian tracers from launch to final deposition in the rotating
Didymos--Dimorphos frame, combining three-dimensional orbital dynamics with
surface interactions on a digital terrain model (DTM) of Dimorphos. This
section first describes the modelling strategy and acceleration field, then
the surface-contact prescription, the dynamic-slope diagnostic, and the
initial conditions used in the simulations.

\subsection{Modelling strategy}
\label{subsec:ravel_kinematic}

\begin{figure}
  \centering
  \includegraphics[width=1\linewidth]{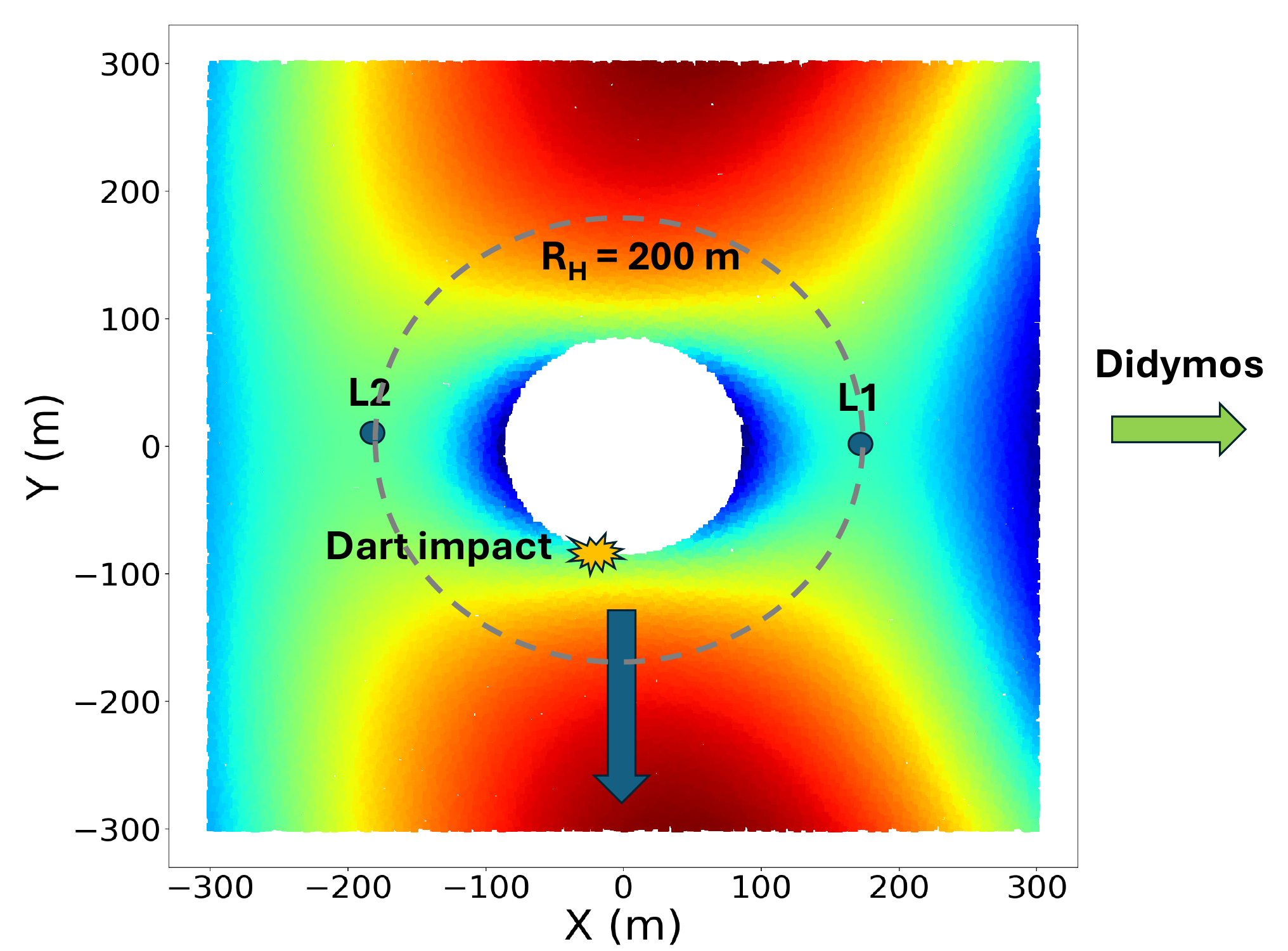}
  \caption{
    Effective potential around Dimorphos, in the XY plane, at the time of
    the DART impact. The white region shows Dimorphos, the yellow star in
    the leading hemisphere marks the DART impact site, L1 and L2 indicate
    the approximate Lagrange points located at a distance of 200~m, and the
    green arrow on the right points towards Didymos.
  }
  \label{fig:potential}
\end{figure}

The dynamical environment of Dimorphos is highly complex. Its assumed synchronous
rotation, pronounced triaxial shape, and proximity to Didymos produce a
strongly asymmetric effective gravity field (Fig.~\ref{fig:potential}),
in which centrifugal, tidal, and Coriolis accelerations are comparable
in magnitude to self-gravity. In this regime, small variations in launch
direction and velocity may lead to markedly different orbital evolution for
low-velocity ejecta, including prompt re-impact, temporary bound motion,
or escape from the immediate vicinity of the body (see Section \ref{sec:results_smooth}).

\begin{figure}
  \centering
  \includegraphics[width=0.8\linewidth]{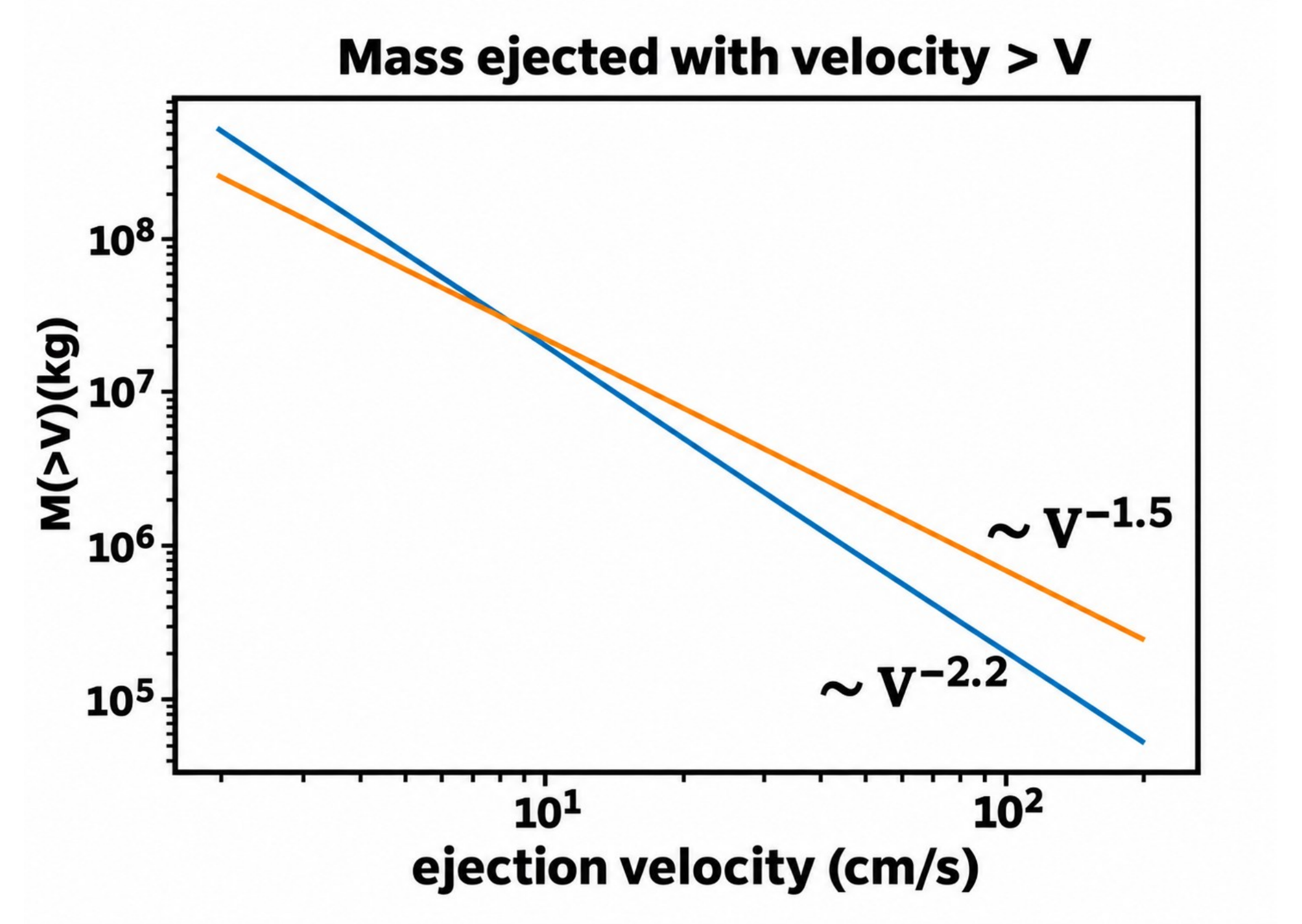}
  \caption{
    Cumulative mass ejected above a given velocity, following a power-law
    relationship $M(>V)\propto V^{-\alpha}$ as summarized by
    \citet{Stickle2025}. The distributions are scaled so that the total
    mass ejected above the Dimorphos escape velocity, about
    $9$~cm\,s$^{-1}$, lies between $1.3$ and
    $6.4\times10^7$~kg, as reported by
    \citet{Graykowski2023,KimJewitt2023,Roth_2023_Alma_DART}.
  }
  \label{fig:mass_ejected_powerlaw}
\end{figure}

Furthermore, the re-accreted ejecta do not necessarily remain near their
landing sites: once on the surface, motion driven by the local effective
gravity field (displayed in Fig.~\ref{fig:surface_accel}) and friction with the surface
can potentially  transport material over distances representing a significant fraction
of the body radius. Figure~\ref{fig:surface_accel} shows that, owing to Dimorphos' slow rotation,
the acceleration field at the surface of Dimorphos may naturally tend to
transport material towards the poles, an effect already identified for slowly
rotating, flattened small bodies by \citet{Dobrovolskis2019}.

Here we use a simple kinematic description aimed at
capturing the large-scale redistribution of ejecta rather than the
grain-scale mechanics of regolith flow. The tracers are massless and have
no intrinsic size; they follow prescribed acceleration fields and are used
as markers of the transport pathways from launch to deposition. This choice
is well suited to the ballistic and orbital phases, where gravity and
rotational accelerations are independent of particle properties. Once a
tracer reaches the surface, its subsequent motion is governed by a reduced
contact model that includes rebound, frictional sliding, and possible
detachment.

\subsection{Dynamical equations and acceleration field}
\label{subsec:ravel_accelerations}

The tracers' motion follows the classic kinematic equations:
\begin{equation}
\label{eq:gov_eqs}
\begin{aligned}
\frac{d\overrightarrow{r}}{dt} &= \overrightarrow{v}, \\
\frac{d\overrightarrow{v}}{dt} &= \overrightarrow{a},
\end{aligned}
\end{equation}
where $\overrightarrow{r}$ is the position vector and
$\overrightarrow{v}$ is the local instantaneous velocity at position
$\overrightarrow{r}$.

All calculations are performed in the Dimorphos body frame, where the
surface is fixed, using Cartesian coordinates. During the ballistic phase,
tracers experience the orbital acceleration
$\overrightarrow{a}_{\rm orb}$, which is the sum of the following terms:
\begin{equation}
\label{eq:orbital_accelerations}
\begin{aligned}
\overrightarrow{a}_{\rm CF} &=
  -\overrightarrow{\omega}_d
  \times (\overrightarrow{\omega}_d \times \overrightarrow{r}) \text{ (Centrifugal acceleration)} \\
\overrightarrow{a}_{\rm COR} &=
  -2 \overrightarrow{\omega}_d \times \overrightarrow{v} \text{ (Coriolis acceleration)}\\
\overrightarrow{a}_{\rm DT} &=
  \frac{-GM_D}{\|\overrightarrow{r}-\overrightarrow{r}_D\|^3}
  (\overrightarrow{r}-\overrightarrow{r}_D)\\
&\quad
  - \frac{GM_D}{\|\overrightarrow{r}_D\|^3}
  \overrightarrow{r}_D \text{ (Didymos tides)}\\
\overrightarrow{a}_{\rm dG} &=
  -\sum_{i}
  \frac{G m_i}{\|\overrightarrow{r}-\overrightarrow{r}_i\|^3}
  (\overrightarrow{r}-\overrightarrow{r}_i) \text{ (Dimorphos self-gravity)}
\end{aligned}
\end{equation}

where $\overrightarrow{r}_D$ is the position vector of Didymos' centroid,
$\overrightarrow{r}_d$ is the position vector of Dimorphos' centroid, and
$i$ is any \textit{mascon} inside Dimorphos.

The gravitational field of Dimorphos is computed using a \textit{mascon
approach}, in which the body is represented by $N_i=10^4$
{equal-mass particles randomly distributed within the Dimorphos surface mesh according to a uniform spatial distribution. Each mascon has a mass
$m_i=M_d/N_i$, where $M_d$ is the total mass of Dimorphos. The value \(N_i=10^4\) was selected as a compromise between computational cost and accuracy}. We adopt
{$M_d = 4.85 \times 10^9$~kg}, assuming that Dimorphos has the same bulk
density as Didymos \citep{Naidu2024}. This corresponds to the classical
\textit{mascon technique} \citep{Scheeres_2010}.

{We tested the numerical convergence of the surface gravity field using
$N_i=10^4$, $10^5$, and $10^6$ mascons, with independent random
realizations of the mascon distribution. The relative differences in the norm of the surface gravity field remain below 15$\%$ {in a few very localized regions,} and below 5$\%$ on average over the { whole }surface. The gravity field progressively converges towards the \(N_i=10^6\) solution as the number of mascons is increased. Because of the computational cost of the full 22~h integrations
with a 1~s timestep, the simulations presented in this work use
$N_i=10^4$ mascons. We verified in a representative case that increasing
the resolution to $N_i=10^5$ does not modify the main dynamical evolution
or the final distribution of the tracers.}

Dimorphos' surface is modelled using a triangular mesh. For the case of a
triaxial ellipsoid, we use one million nodes with radii $a = 86.5$~m,
$b = 85$~m, $c = 56.5$~m following \citet{Daly_2024}, with the long axis
pointing to Didymos when Dimorphos is at its pericenter. {For the DART/DRACO-based shape model, we use the updated Dimorphos shape model of \citet{Daly_2024}}. Detection of
contact of the tracers with the mesh, and calculation of local surface normals for each facet are done using the Python library \textsc{PyVista}
\citep{Pyvista_sullivan2019pyvista}.

\subsection{Surface contact, rebound, and sliding}
\label{subsec:surface_contact}

 {The detailed surface treatment is described in \cite{Herreros_Charnoz_Phobos_2026} whose main aspects are summarized herein.} When a tracer is in contact with the surface, its motion is affected by
the orbital acceleration field, $\overrightarrow{a}_{\rm orb}$, together
with two additional contact-related acceleration terms: a normal reaction
acceleration, $\overrightarrow{a}_{\rm R}$, and a frictional acceleration,
$\overrightarrow{a}_{\rm F}$. The total acceleration is therefore written
as
\begin{equation}
\label{eq:total_surface_acceleration}
\overrightarrow{a} =
\overrightarrow{a}_{\rm orb}
+ \overrightarrow{a}_{\rm R}
+ \overrightarrow{a}_{\rm F}.
\end{equation}

The normal reaction acceleration is directed along the outward local normal
$\overrightarrow{n}$ to the Dimorphos surface at the contact point,
\begin{equation}
\label{eq:normal_reaction_acceleration}
\overrightarrow{a}_{\rm R} = R \overrightarrow{n},
\end{equation}
where
\begin{equation}
\label{eq:normal_reaction_coefficient}
R = \max\left(0, -\overrightarrow{a}_{\rm orb}\cdot\overrightarrow{n}\right).
\end{equation}
Here, $R$ has units of acceleration and represents the normal reaction per
unit mass. This term is only applied when the orbital acceleration has a
component directed into the surface. If this condition is not satisfied,
the reaction vanishes and the tracer is allowed to detach from the surface.

The frictional acceleration is modelled using a Coulomb-type law and acts
opposite to the tangential direction of motion along the surface. Defining
\begin{equation}
\label{eq:tangential_velocity}
\overrightarrow{v}_{\parallel} =
\overrightarrow{v}
-
\left(
\overrightarrow{v}\cdot\overrightarrow{n}
\right)
\overrightarrow{n}
\end{equation}
as the tangential velocity, and
\begin{equation}
\label{eq:tangential_unit_vector}
\overrightarrow{u}_{\parallel} =
\frac{\overrightarrow{v}_{\parallel}}
{|\overrightarrow{v}_{\parallel}|+\epsilon}
\end{equation}
as the regularized tangential direction of motion, where
$\epsilon = 10^{-9}\,\mathrm{m\,s^{-1}}$ is introduced to avoid
numerical singularities as the tangential velocity approaches zero.
The frictional acceleration is then written as
\begin{equation}
\label{eq:frictional_acceleration}
\overrightarrow{a}_{\rm F} =
-\tan\phi \, R \, \overrightarrow{u}_{\parallel},
\end{equation}
where $\phi$ is the effective friction angle.

 {As described in \citet{Herreros_Charnoz_Phobos_2026},
at very low tangential velocities, the friction law switches
from the dynamic to the static regime.
In this regime, a static frictional acceleration
$\overrightarrow{a}_{\rm S}$, parallel to the local surface,
balances the other tangential accelerations to keep the
particle at rest, provided that}
\begin{equation}
\label{eq:static_frictional_acceleration}
|\overrightarrow{a}_{\rm S}| \leq \tan\phi\,R.
\end{equation}
 {If this condition is not satisfied, the particle resumes sliding. We assume the same friction angle, \(\phi\), for both the dynamic and static regimes (Eqs. \ref{eq:frictional_acceleration} and \ref{eq:static_frictional_acceleration}, respectively), following \citet{Herreros_Charnoz_Phobos_2026}.}

 {
When a tracer is in contact with the surface, the velocity component
directed into the surface is removed at each time step, as described in
\cite{Herreros_Charnoz_Phobos_2026}.
Since the surface is represented by a triangulated polyhedron,
each facet is locally planar, and no explicit curvature term is
introduced in the normal reaction.
Instead, the effects of surface curvature are accounted for through
successive changes in the local surface normal and the corresponding
velocity corrections as particles move across adjacent facets.
Importantly, tracers are not constrained to remain on the surface
and can detach whenever their motion becomes outward.
We validated the numerical treatment of surface curvature and particle detachment using the classical problem of frictionless sliding on a sphere under uniform gravity (Appendix~\ref{annex_section_deteachement_sphere}), obtaining good convergence towards the analytical solution.
}

In the simulations presented below, rebounds are treated as inelastic with
a normal coefficient of restitution $\epsilon_n = 0.1$, and sliding is
computed with $\phi = 35^\circ$. The adopted friction angle of $\phi=35^\circ$ is consistent
with the bulk internal friction angle of $32.7\pm2.5^\circ$
inferred from the morphological analysis of Dimorphos
surface boulders by \citet{robin2024}. 

{When a tracer impacts the surface, an inelastic rebound is first applied
to the normal component of its velocity:
\begin{equation}
v_n^{+}=-\epsilon_n v_n^{-},
\end{equation}
where $v_n^{-}$ and $v_n^{+}$ denote the normal velocity components
immediately before and after impact, respectively. If the tracer remains in contact with the surface after the collision, its subsequent motion is governed by the contact and friction formulation of Eqs.~(\ref{eq:total_surface_acceleration})--(\ref{eq:frictional_acceleration}).}

This formulation does not force Lagrangian tracers to remain attached to
the surface. When the orbital acceleration has no component directed into
the surface, the normal reaction is set to zero and the tracer is allowed
to detach.

The time integration scheme used to solve the system of ODEs is a simple
first-order Euler solver with a time-step of $\Delta t = 1$~s, which is
much smaller than the orbital period of the binary system ($\sim 11$~h).

\begin{figure}
  \centering
  %\includegraphics[width=1\linewidth]{IMAGES/Fig2_SM.png}
%\includesvg[width=0.95\columnwidth]{IMAGES/NEW_FIGURES/Fig3} 

 \includegraphics[width=1\linewidth]{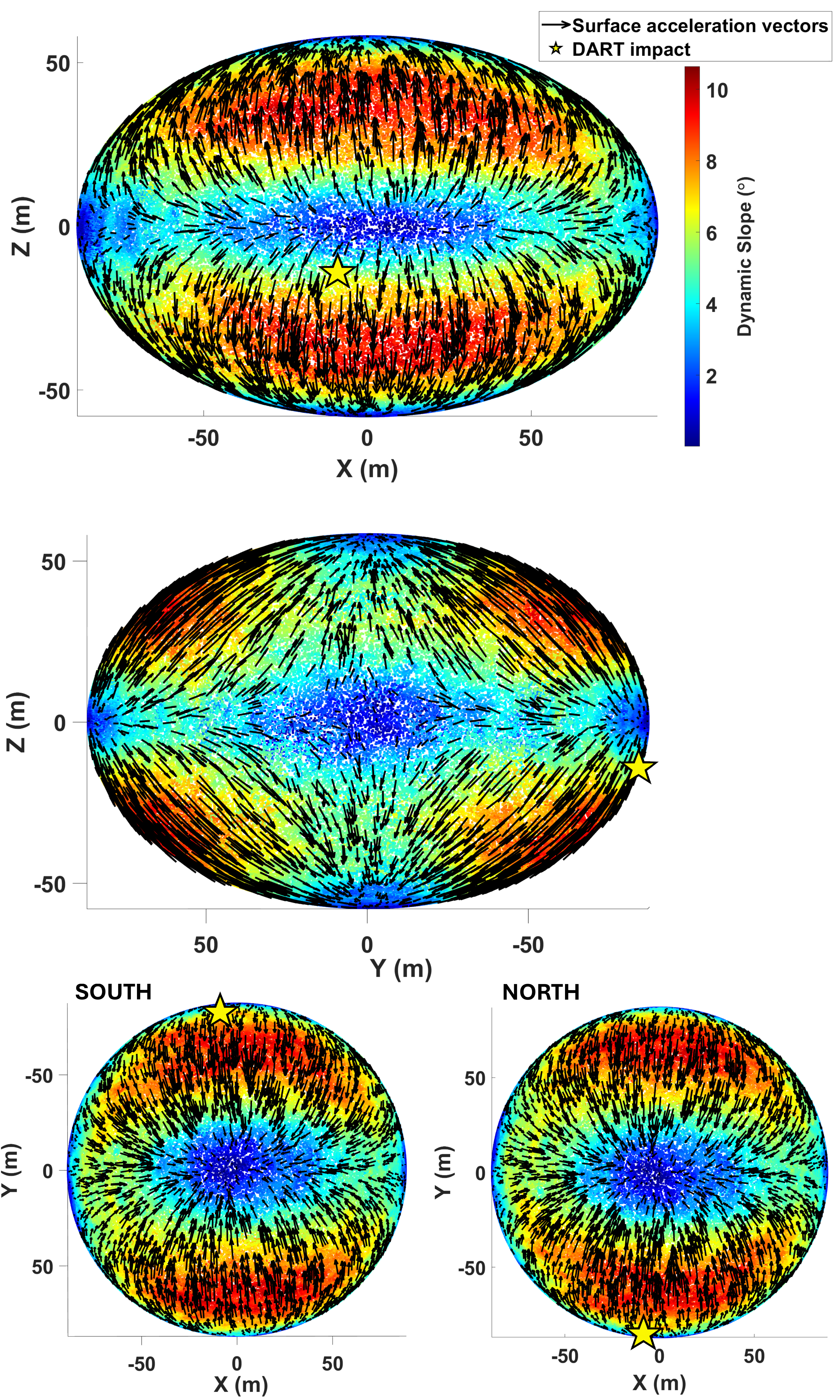}
 
  \caption{
    Dynamic slope and surface effective acceleration field on the smooth
    ellipsoidal model of Dimorphos. Colours represent the dynamic slope,
    while black arrows show the tangential component of the effective
    acceleration projected onto the surface. The effective acceleration
    accounts for Dimorphos self-gravity, centrifugal acceleration, and
    Didymos tidal acceleration. The yellow star marks the DART impact site.
    The upper panel shows the XZ projection, and the lower panels show the
    south and north polar projections.
  }
  \label{fig:surface_accel}
\end{figure}

\subsection{Surface dynamic slope}
\label{subsec:ravel_dynamic_slope}

Let $\overrightarrow{a}_{\text{eff}}$ be the effective local acceleration
at the surface (including gravitational, centrifugal and tidal components)
(Fig.~\ref{fig:surface_accel}), and let $\overrightarrow{n}$ denote the
outward unit normal to the local surface. The angle $\theta$ between
$\overrightarrow{a}_{\text{eff}}$ and $\overrightarrow{n}$ is defined
through
\begin{equation}
\label{eq:dynamic_slope_cosine}
\cos \theta = \frac{\overrightarrow{a}_{\text{eff}} \cdot
                 \overrightarrow{n}}
                {\|\overrightarrow{a}_{\text{eff}}\| \,
                 \|\overrightarrow{n}\|}.
\end{equation}
Since \(\overrightarrow{n}\) is a unit vector, the local \textit{dynamic slope} (Fig.~\ref{fig:surface_accel}) can be written as
\begin{equation}
\label{eq:dyn_slope}
\theta = \pi - \arccos\left(
  \frac{\overrightarrow{a}_{\text{eff}} \cdot \overrightarrow{n}}
       {\|\overrightarrow{a}_{\text{eff}}\|}
\right),
\end{equation}
which is 0, by convention, when the acceleration is perpendicular to the
surface and pointing downward.

\subsection{Numerical setup and initial conditions}
\label{subsec:initial_conditions}

In our standard case, Dimorphos is represented by a triaxial ellipsoid with
semi-axes $a = 86.5$~m, $b = 85$~m, and $c = 56.5$~m
\citep{Daly_2024}, discretized with $10^6$ surface points. The masses of
Didymos and Dimorphos are $5.26 \times 10^{11}$~kg {and
$4.85 \times 10^{9}$~kg}, respectively, assuming an average density of
$2790\,\mathrm{kg\,m^{-3}}$ \citep{Naidu2024}.
Dimorphos is placed on its post-impact orbit, with orbital period
11.367~h and eccentricity $e = 0.0274$ \citep{Chabot2025}, and its motion
and rotation are integrated together with the tracers.

{The crater radius remains uncertain, with hydrocode-based estimates of $\sim 10$--30~m \citep{Stickle2025}, while the excavation extent inferred from ejecta-cone modelling may reach several tens of metres \citep{Hirabayashi_2025}. We therefore adopt $R_c=35$~m for the spatial extent of the ejecta source as a fiducial intermediate value between these estimates. We also consider a smaller $R_c=15$~m case to assess the sensitivity of our results to the assumed spatial extent of the ejecta source region (Sect.~\ref{sec:radius_15m}).}

The ejecta mass--velocity distribution follows \citet{Stickle2025}. We adopt as a fiducial case
\begin{equation}
\label{eq:mass_velocity}
M(>v) \propto v^{-2}.
\end{equation}
{within the range of slopes illustrated in Fig. \ref{fig:mass_ejected_powerlaw}.
 {The normalization is chosen such that the mass ejected above the escape
velocity of Dimorphos ($9$~cm~s$^{-1}$) is $3.8 \times 10^7$~kg,
 as an average of the observational estimates of \cite{Graykowski2023,KimJewitt2023, Roth_2023_Alma_DART}  (1.3-6.4 $\times$ 10$^7$ kg), which differ by approximately a factor of five.
We adopt the $v^{-2}$ scaling reported by \citet{Stickle2025},
although its validity at very low velocities remains unconstrained
by impact simulations.
Extrapolating this power law towards lower velocities can yield a total
mobilized mass comparable to, or exceeding, the mass of Dimorphos,
highlighting the limitations of such an extrapolation.
However, the slowest ejecta primarily undergo short-range transport
and redeposition rather than escaping the asteroid.
Consequently, the adopted scaling should be regarded as a prescription
for the relative weighting of velocity populations, rather than a
reliable estimate of their absolute masses.
Our conclusions regarding velocity-dependent transport remain unaffected
by this normalization, whereas the absolute mass budget remains uncertain.}
Since tracers are sampled uniformly in
velocity, the corresponding differential mass distribution is $-dM/dv\propto v^{-3}$,
and individual tracers are therefore weighted as $m(v)\propto v^{-3}$.} 

\begin{figure}
  \centering
  \includegraphics[width=1.0\linewidth]{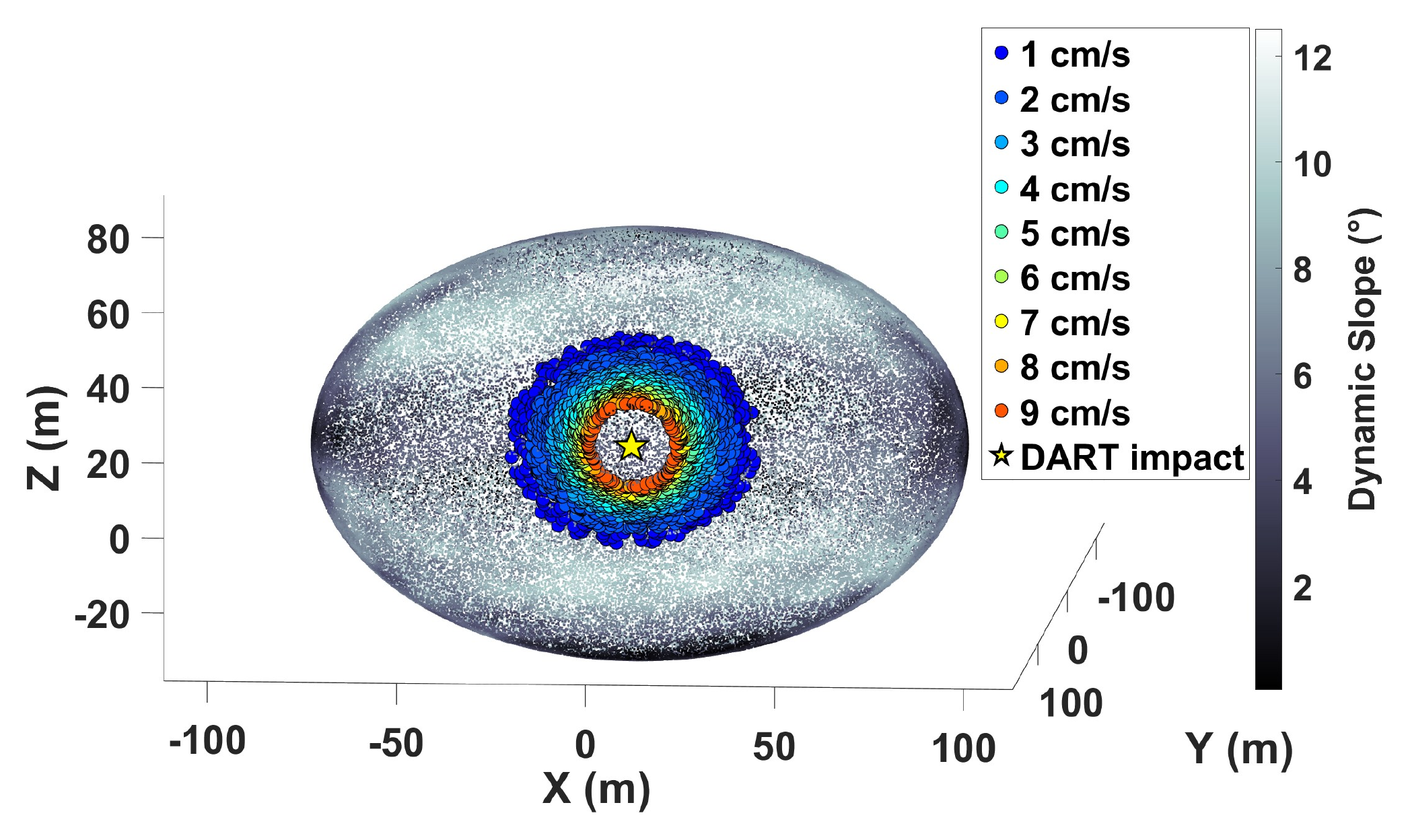}
  %\includesvg[width=1\columnwidth]{IMAGES/NEW_FIGURES/Fig4} 
  \caption{
    Initial conditions for the standard smooth ellipsoidal Dimorphos case
    with $R_c = 35$~m. Coloured points show the initial tracer positions
    around the DART impact site, with colour indicating the initial
    ejection velocity $v_0$. The grey-scale background represents the
    dynamic slope on the surface, and the yellow star marks the DART impact
    site.
  }
  \label{fig:initial_velocities}
\end{figure}

{The ejection angle $\theta$, measured with respect to the plane perpendicular
to the ejection-cone axis, is observationally motivated by the range
$23.5^\circ<\theta<42.5^\circ$ inferred from the DART ejecta plume by
\citet{Hirabayashi_2025}. These observations do not provide a direct
constraint on the launch-angle distribution of the specific low-velocity
ejecta population considered here. In the absence of such a constraint, we
adopt this observationally motivated range rather than introducing an
independent angular distribution. Each simulation therefore uses three
batches of 4000 tracers with $\theta=23.5^\circ$, $33^\circ$, and
$42.5^\circ$, and initial velocities uniformly sampled between
$1$ and $9\,\mathrm{cm\,s^{-1}}$. Initial tracer positions are distributed concentrically around the DART
impact site (Fig.~\ref{fig:initial_velocities}) according to the
Housen--Holsapple scaling \citep{Housen_Holsapple_1983},
\begin{equation}
\label{eq:Housen_Holsapple_law}
r(v) =
R_c\left(
\frac{v}{1\,\mathrm{cm\,s^{-1}}}
\right)^{-1/2}.
\end{equation}
This relation corresponds to $v\propto r^{-2}$, or equivalently
$\mu=0.5$ in the Housen--Holsapple ejecta scaling, within the experimentally
determined range $0.35\lesssim\mu\lesssim0.55$
\citep{Housen_Holsapple_1983,Housen_Holsapple_2011}. For the velocity range
considered here, $1\leq v\leq9\,\mathrm{cm\,s^{-1}}$,
Eq.~(\ref{eq:Housen_Holsapple_law}) places the launch positions within an
annulus extending from
$R_{\min}=R_c/3$ to $R_c$.}

\subsection{Assumptions and limitations}
\label{subsec:ravel_limitations}

The model is intentionally simplified. Tracers are massless, and
grain-scale friction, cohesion, fragmentation, collisions, and
size-dependent interactions are not resolved. The results should therefore
be interpreted as a first-order dynamical mapping of where low-velocity
ejecta re-impact, migrate, and accumulate, rather than as a grain-scale
simulation of regolith flow. {The surface-interaction model is likewise
intended to capture the first-order effects of rebound and frictional
sliding, rather than to provide quantitative constraints on the mechanical
properties of the regolith. Within these limits, the predicted near-crater
blanket, distal antipodal and trailing deposits, and topographically channelled depositional structures
provide qualitative, testable signatures of the coupled effects of ejecta velocity,
binary dynamics, and  surface topography.}

%======================================================================
\section{Results for the standard case: a smooth ellipsoidal model
         of Dimorphos}
\label{sec:results_smooth}

\begin{figure*}
  \centering
  \includegraphics[scale=0.35]{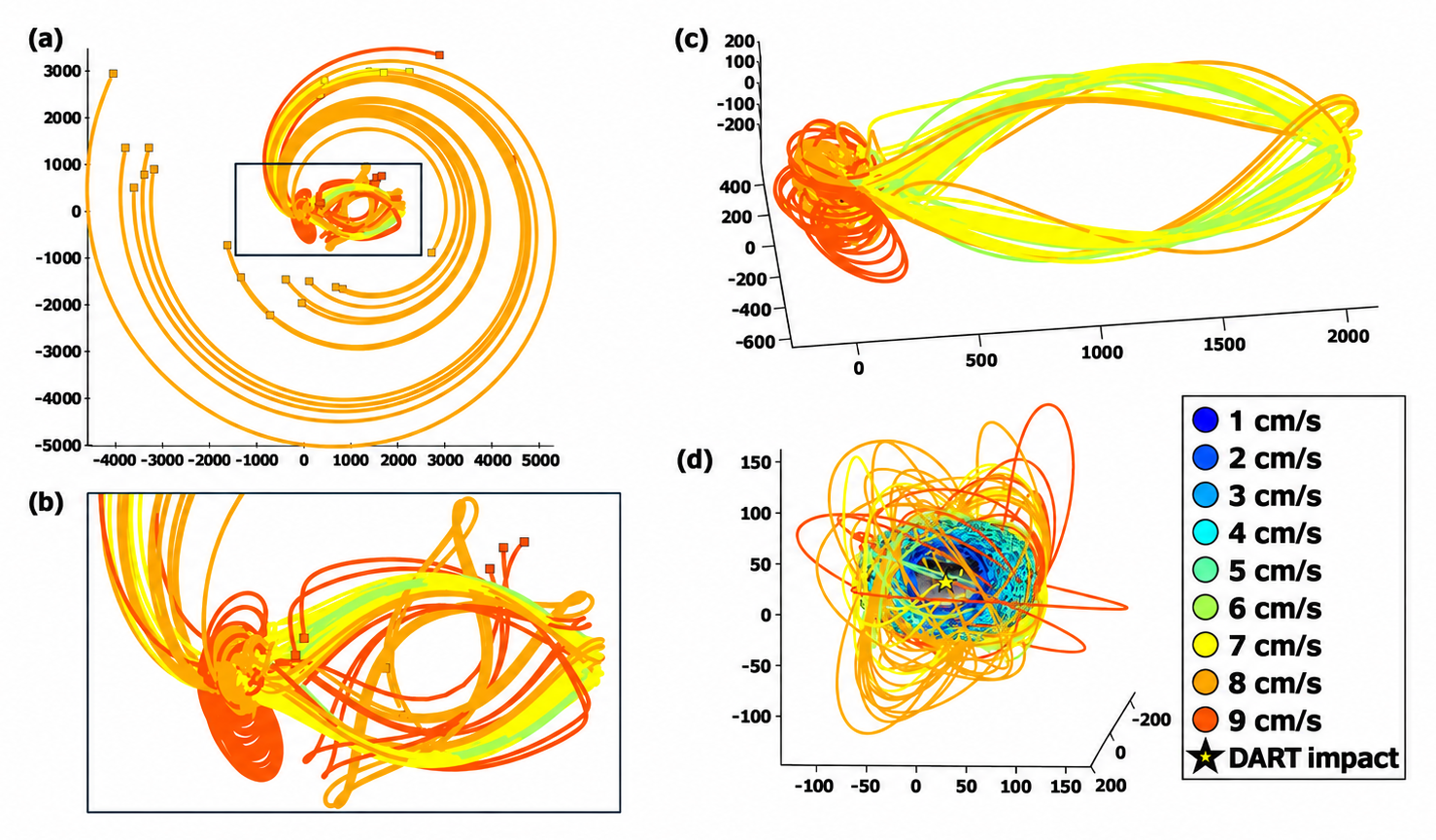}
  \caption{
    Trajectories of the fastest ejecta in the frame co-rotating with
    Dimorphos during the first 22~h after the DART impact.
    Colours indicate initial ejection velocity.
    (a) All trajectories, including those escaping the system.
    (b) Trajectories trapped in the binary system; although not shown,
        Didymos lies near the centre of the main loop, so many
        high-velocity trajectories may re-impact it.
    (c) Trajectories re-impacting Dimorphos after filtering out the ones
        impacting Didymos.
    (d) Trajectories around Dimorphos after first contact, including
        rebounds and surface motion.
  }
  \label{fig:orbits}
\end{figure*}

During the first 22~h, ejecta trajectories fall into four regimes. A small
fraction of tracers ($<1$\%) do not re-impact Dimorphos, either because
they escape the system or because they impact Didymos
(Fig.~\ref{fig:orbits}a,b). Some fast tracers, with initial ejection
velocities $v > 6$~cm\,s$^{-1}$, circulate around Didymos before
re-accreting onto Dimorphos after $\sim 10$--11~h
(Fig.~\ref{fig:histogram_impact_times_Nparticles}a,b), with re-impact
locations distributed over much of the surface
(Fig.~\ref{fig:orbits}c). Other tracers remain inside Dimorphos' Hill
sphere and return to the surface after a partial orbit, often on the
opposite hemisphere (Fig.~\ref{fig:orbits}c); this behaviour can occur for
$v > 3$~cm\,s$^{-1}$ and even for some tracers launched at
9~cm\,s$^{-1}$, because the gravitational perturbation of Didymos can keep
part of this population bound to the binary environment, driving rapid
re-accretion after partial orbits. Finally, tracers with
$v \lesssim 4\,\mathrm{cm\,s^{-1}}$ re-impact quickly near their launch
site (Fig.~\ref{fig:orbits}d).

The trajectory field is strongly anisotropic because Didymos' tides enhance
transport through the Lagrange $L_1$ region
(Fig.~\ref{fig:potential}). This controls the timing of re-impact
(Fig.~\ref{fig:histogram_impact_times_Nparticles}a): more than 95\% of
tracers re-impact on Dimorphos within 5~h, well before Dimorphos completes
one orbit around Didymos ($\sim 11$~h). The first $\sim 10$~h are dominated
by a decreasing flux from short-range trajectories; after $\sim 11$~h, a
second delayed phase begins as tracers that circulated around Didymos
return. By 20~h, more than 99.9\% of the ejected tracers have re-accreted.
{When tracers are weighted according to our fiducial mass--velocity distribution (Eq.~\ref{eq:mass_velocity}), 99.9\% of the low-velocity mass re-accretes within 5~h. This shorter mass-weighted timescale reflects the concentration of mass in the slowest ejecta, which dynamically re-impact most rapidly.}

\subsection{Distribution of first-contact impacts}
\label{subsec:first_contact}

Because RAVEL follows both orbital evolution and surface motion, we
distinguish first-contact locations from final resting positions. Between
them, material may bounce and then slide under the combined effect of local
slopes and friction. Since most impacts occur at $< 8\,\mathrm{cm\,s^{-1}}$,
catastrophic fragmentation and secondary ejecta production during
sesquinary impacts are not expected to be dominant and are not included.

\begin{figure}
  \centering
  \includegraphics[width=1\linewidth]{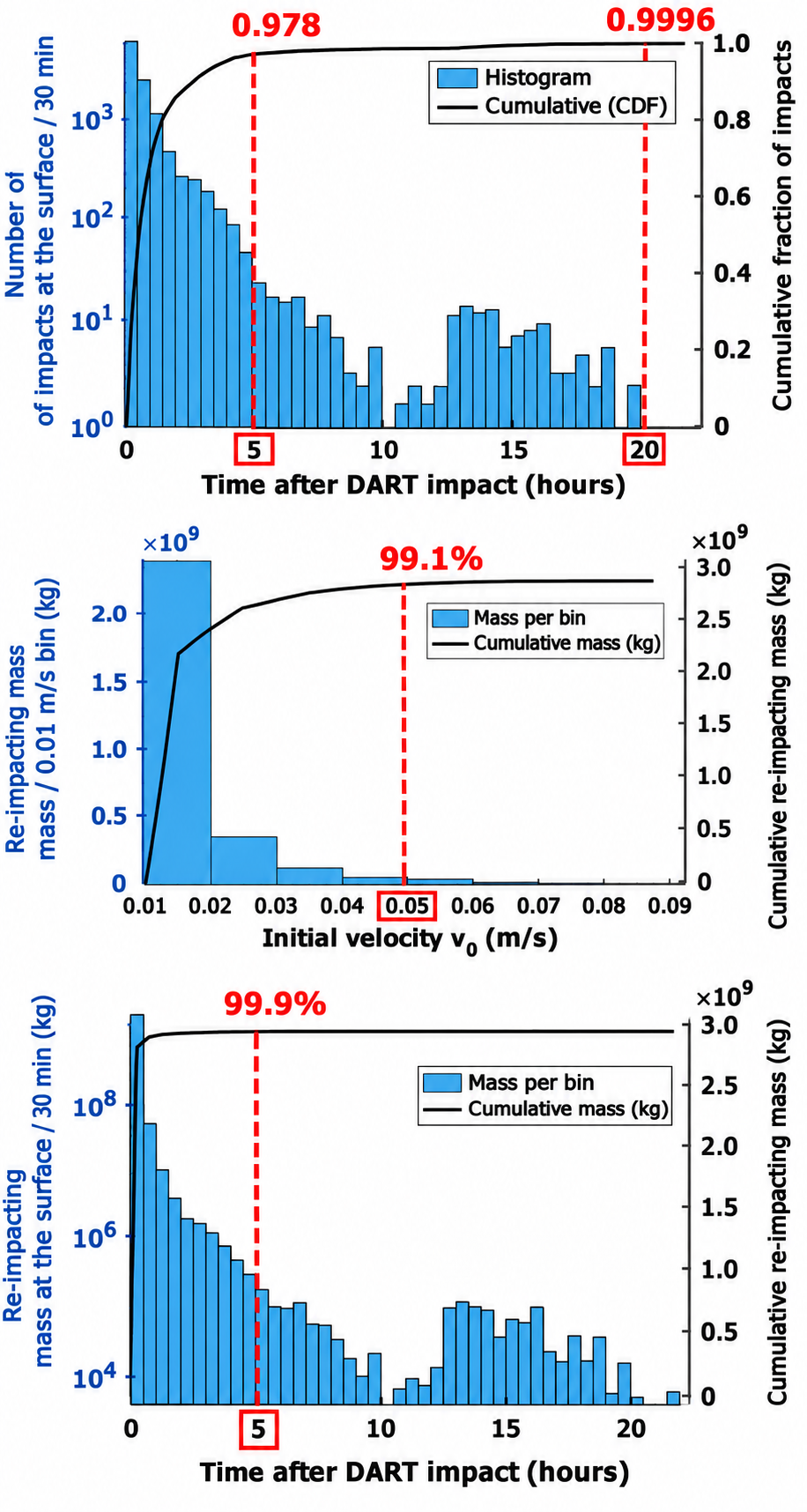}
  
  %\includesvg[width=0.6\columnwidth]{IMAGES/NEW_FIGURES/Fig6} 
  \caption{
    Timing and mass distribution of re-accreting ejecta in the standard
    smooth ellipsoidal case.
    (a) Histogram and cumulative fraction of first-contact times.
    (b) Mass distribution versus initial ejection velocity from the
        adopted scaling law given in Eq.~(\ref{eq:mass_velocity}).
    (c) Mass-weighted first-contact times, calculated using 
      Eq.~(\ref{eq:mass_velocity}).
  }
  \label{fig:histogram_impact_times_Nparticles}
\end{figure}

First-contact locations are shown in Fig.~\ref{fig:figure_smooth_dimorphos}a
in East-longitude/North-latitude projection. They correlate strongly with
ejection velocity: tracers with $v \lesssim 4$~cm\,s$^{-1}$ re-impact near
the DART impact site ($264.3^\circ$E, $-8.8^\circ$N;
\citealt{Chabot_2024}), over about half of the impacted hemisphere, whereas
tracers with $v \gtrsim 7$~cm\,s$^{-1}$ generally reach the opposite
hemisphere. This produces an antipodal concentration, while the polar
regions are less affected. The pattern for $v \gtrsim 6$~cm\,s$^{-1}$
agrees with the long-term results of \citet{Langner_2024}, but here it
appears within the first $\sim 20$~h, when most of the mass is re-accreted
(Fig.~\ref{fig:histogram_impact_times_Nparticles}c).

\subsection{Rebounds, sliding, and final surface distribution}
\label{subsec:rebounds_sliding}

\begin{figure*}[t]
  \centering

      \begin{minipage}[t]{0.5\textwidth}
        \centering

  \includegraphics[width=\linewidth]{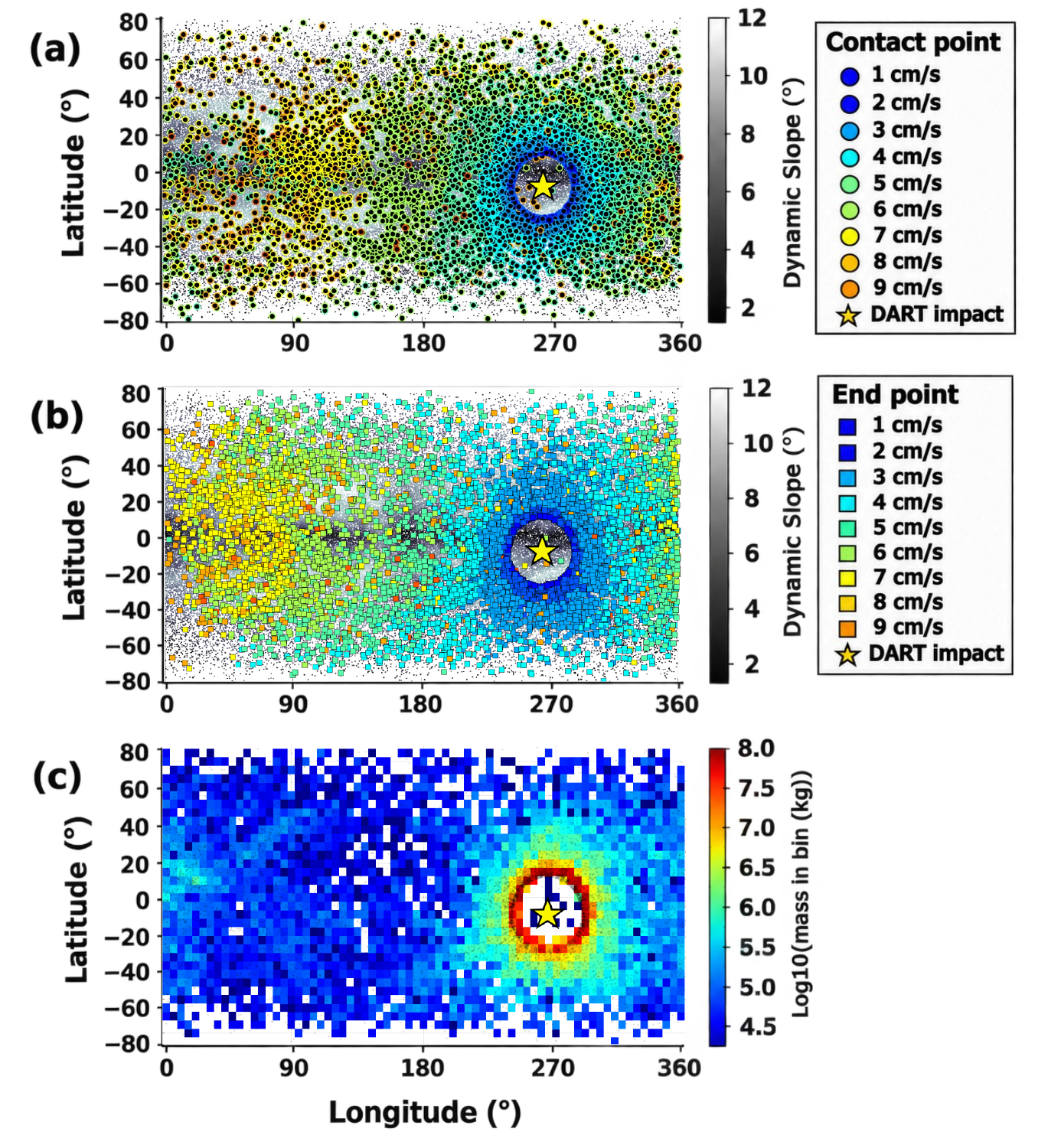}
  \caption{
    Distribution of ejecta on the standard smooth ellipsoidal model
    ($R_c = 35$~m; ejection angles with respect to the plane perpendicular
    to the ejection-cone axis: $23.5^\circ$--$42.5^\circ$).
    (a) First-contact locations.
    (b) Final resting positions.
    (c) Mass-weighted final distribution, expressed as $\log_{10}$ of the
        mass per bin ($5^\circ \times 5^\circ$).{ White regions correspond to surface cells containing no deposited tracers at the end of the simulation and therefore represent zero deposited mass in our tracer sampling.}
    Slow ejecta remain near the impact site, faster tracers populate distal
    and antipodal regions, and the mass-weighted map is dominated by the
    slowest ejecta.
  }
  \label{fig:figure_smooth_dimorphos}
    \end{minipage}
    \hfill
    \begin{minipage}[t]{0.49\textwidth}
        \centering
  \includegraphics[width=\linewidth]{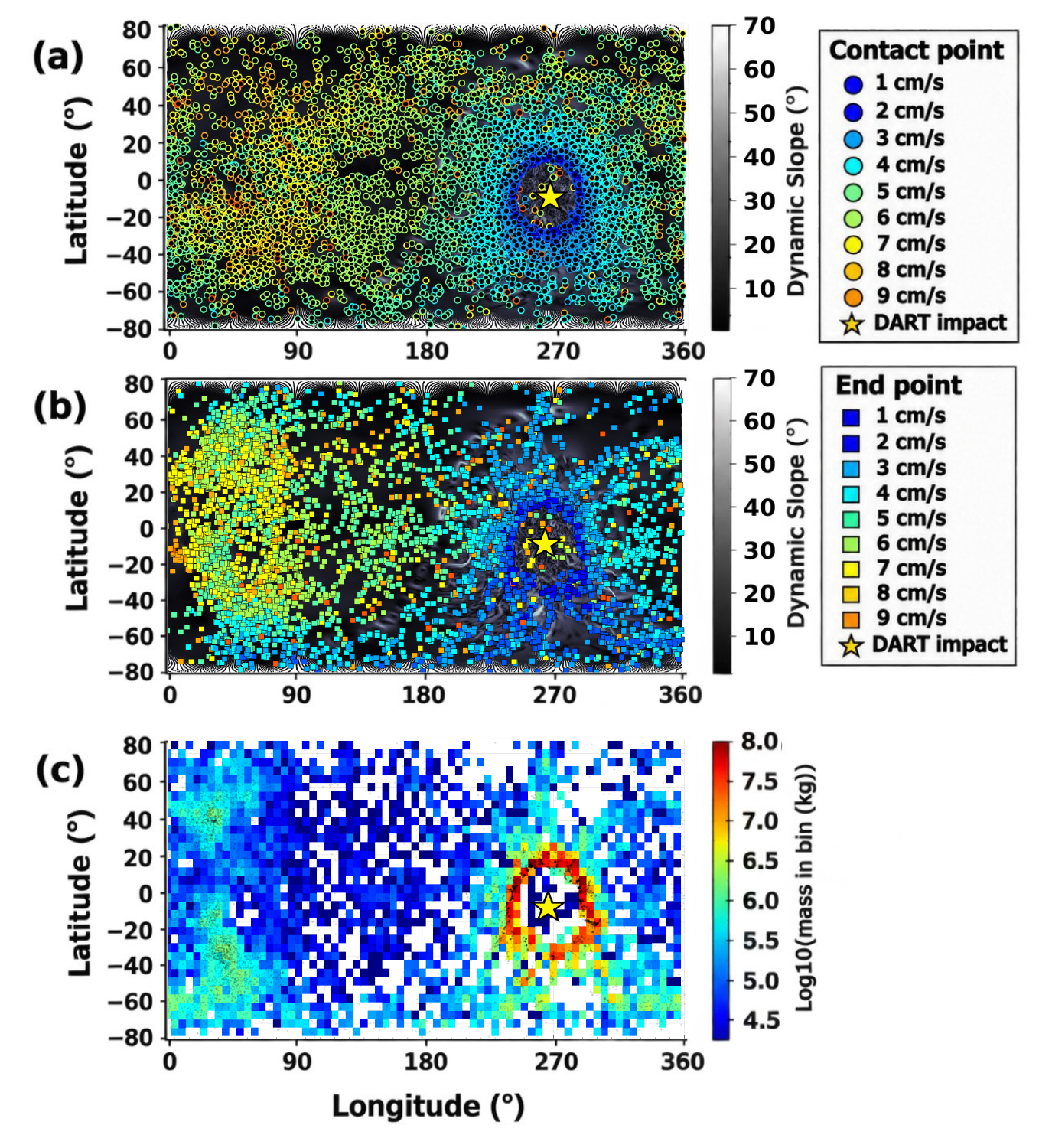}
  \caption{
    Distribution of ejecta on the DART/DRACO-based DTM (DRDTM)
    ($R_c = 35$~m; ejection angles: $23.5^\circ$--$42.5^\circ$).
    (a) First-contact locations.
    (b) Final resting positions.
    (c) Mass-weighted final distribution, expressed as $\log_{10}$ of the
        mass per bin ($5^\circ \times 5^\circ$).{ White regions correspond to
surface cells containing no deposited tracers at the end of the simulation and therefore represent zero deposited mass in our tracer
sampling.}
    The roughness around the impact region produces ray-like near-crater
    deposits dominated by the slowest ejecta; the distal fast-ejecta
    accumulation remains visible although the non-imaged hemisphere is
    smooth.
  }
  \label{fig:figure_DART_DTM_impacts}
    \end{minipage}

\end{figure*}

After first contact (displayed in Fig.~\ref{fig:figure_smooth_dimorphos}a),
tracers may rebound or transition into frictional sliding. We adopt a
normal restitution coefficient of 0.1, consistent with low-velocity impacts
into regolith simulants
\citep{Colwell_1999,Brisset_2019,Brisset_2020}. Tracers with
$v \gtrsim 6$~cm\,s$^{-1}$ typically rebound and follow long ballistic
paths, sometimes spanning one hemisphere before settling, while
lower-velocity tracers mainly undergo short-range frictional surface
motion. Most bouncing tracers experience only one or two rebounds before
sliding to rest.

\begin{figure}
  \centering

\includegraphics[width=1\linewidth]{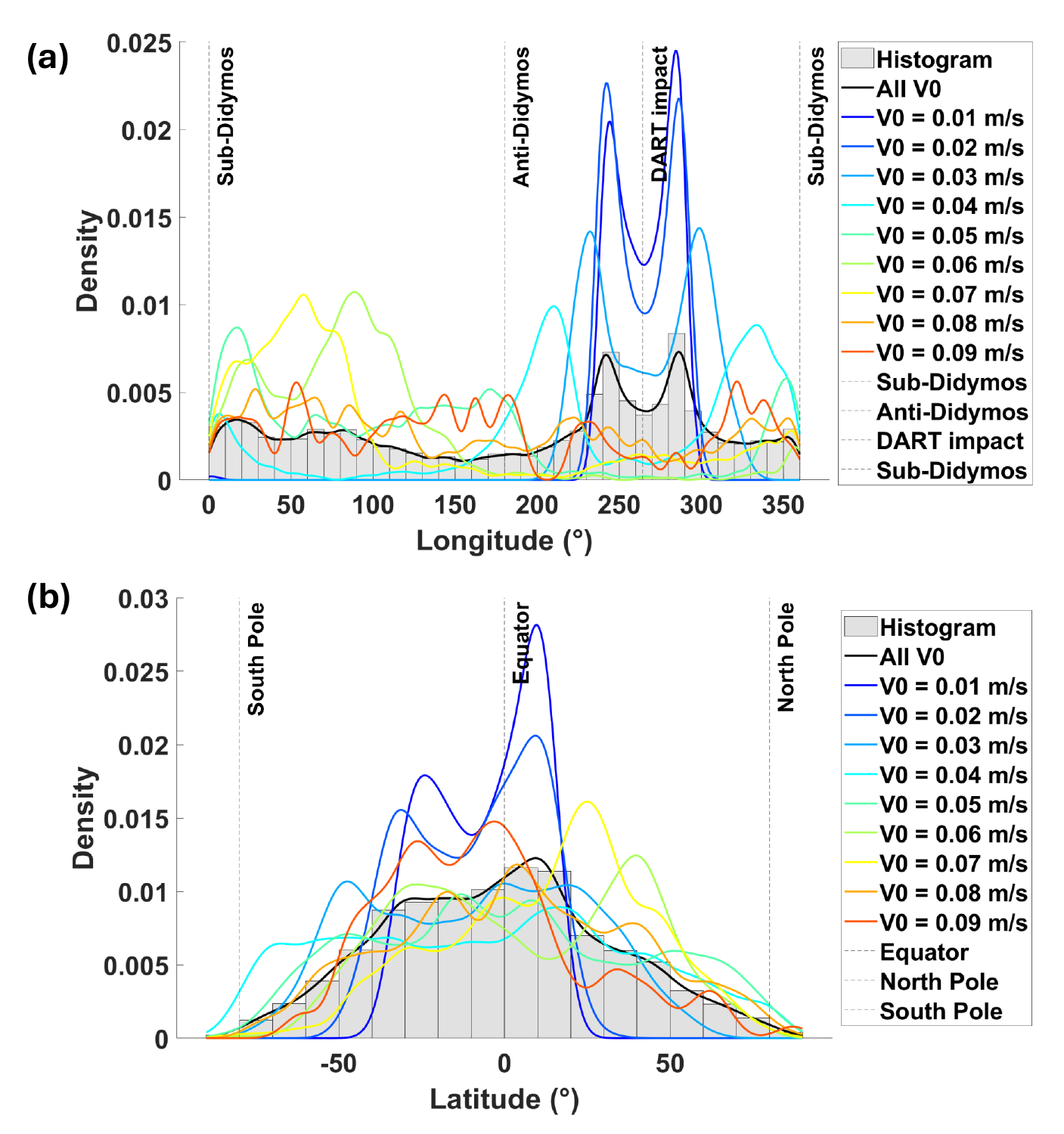}
  %\includesvg[width=1\columnwidth]{IMAGES/NEW_FIGURES/Fig7} 
  \caption{
    Longitude and latitude distributions of the final tracer positions for
    the standard smooth ellipsoidal Dimorphos case with $R_c = 35$~m.
    (a) Distribution as a function of longitude.
    (b) Distribution as a function of latitude.
    Grey bars show the normalized histograms of all final positions, and
    black curves show kernel density estimates (KDEs) computed from all
    end-points. Coloured curves show KDEs computed separately for each
    initial ejection velocity. Because the velocity-resolved KDEs are
    independently normalized, they represent conditional distributions for
    each velocity class rather than additive contributions to the total
    density.
  }
  \label{fig:histogram_long-lat_smooth_35m}
\end{figure}

Figure~\ref{fig:figure_smooth_dimorphos}b shows the final ejecta
distribution after rebound and frictional sliding. Despite the relatively
high friction angle ($35^\circ$), {surface transport after first contact}
substantially modifies the first-contact pattern
(Fig.~\ref{fig:figure_smooth_dimorphos}a). Even relatively slow ejecta launched at
$\sim 4$~cm\,s$^{-1}$ migrate by $\sim 40^\circ$--$50^\circ$ in latitude
and longitude (compare Figs.~\ref{fig:histogram_long-lat_smooth_35m} and \ref{fig:figure_smooth_dimorphos}a,b), while
particles with $v \gtrsim 7$~cm\,s$^{-1}$ can be displaced by nearly half
a hemisphere relative to their initial re-impact location. After first
contact, trajectories typically extend over $\sim 20$--40\% of the body
radius and are strongly curved by Coriolis forces in the rotating frame.
The resulting deposit covers most of Dimorphos' surface but is markedly
heterogeneous: slow ejecta ($\lesssim 5$~cm\,s$^{-1}$) remain concentrated
around the DART impact site, whereas faster material ($\gtrsim 7$~cm\,s$^{-1}$)
preferentially accumulates in the antipodal and trailing hemispheres,
leaving depleted regions near longitudes $\pm 90^\circ$ from the impact
point. The smooth case therefore reveals a strong velocity sorting of the
final ejecta blanket.

%======================================================================

%--------- figs ----------------

\begin{figure}
  \centering
  \includegraphics[width=0.75\linewidth]{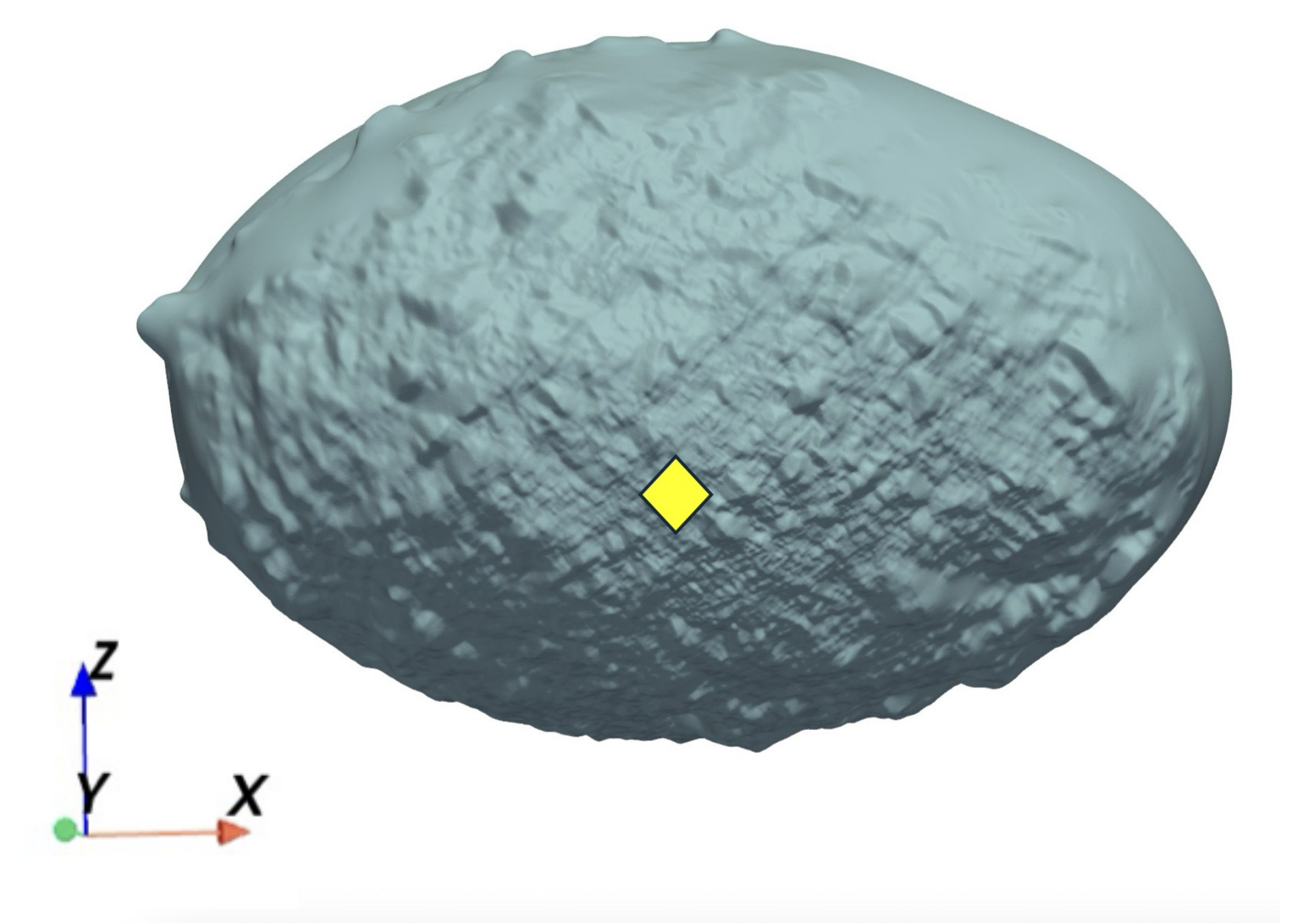}
  \caption{
    DART/DRACO-based Dimorphos terrain model (DRDTM). The yellow diamond
    marks the DART impact site. The imaged hemisphere contains the
    resolved topography used in the DRDTM simulations, whereas the opposite
    hemisphere remains comparatively smooth because it was not imaged by
    DART/DRACO.}
  \label{fig:DART_DTM}
\end{figure}

\begin{figure}
    \centering
    \includegraphics[width=1
    \linewidth]{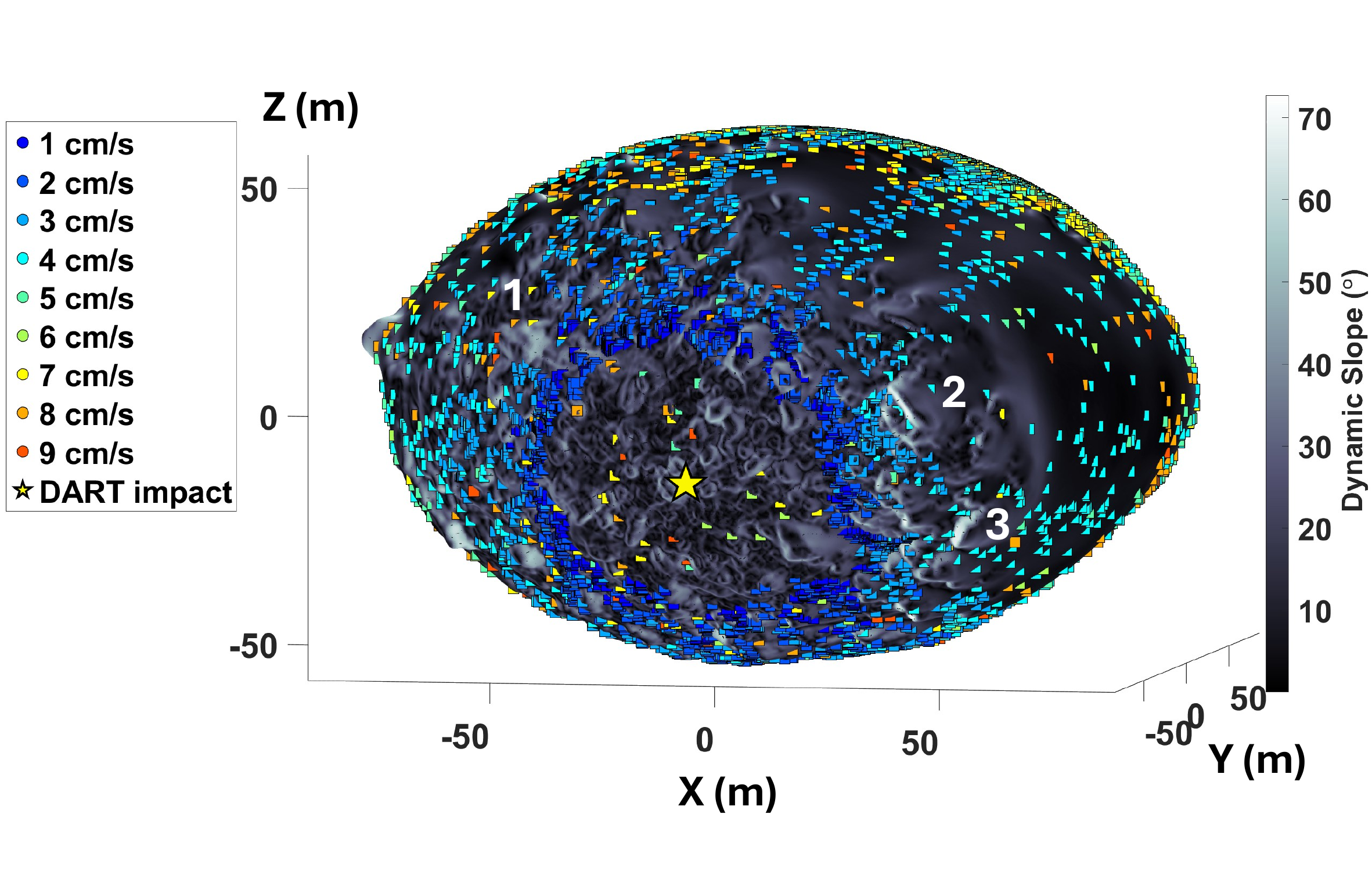}
    \caption{ Three-dimensional view of the final ejecta distribution on the
DRDTM for $R_c=35$~m. Colours indicate initial ejection velocity
and grey levels the local dynamic slope.
Labels 1--3 highlight topographic features associated with
ejecta channelling (streaks) and shadowing (depleted regions).
These mechanisms are examined in Appendix~\ref{annexe_topographic_effects}.
}
    \label{fig:3D_dart_model}
\end{figure}

\begin{figure}
  \centering
  \includegraphics[scale=0.19]{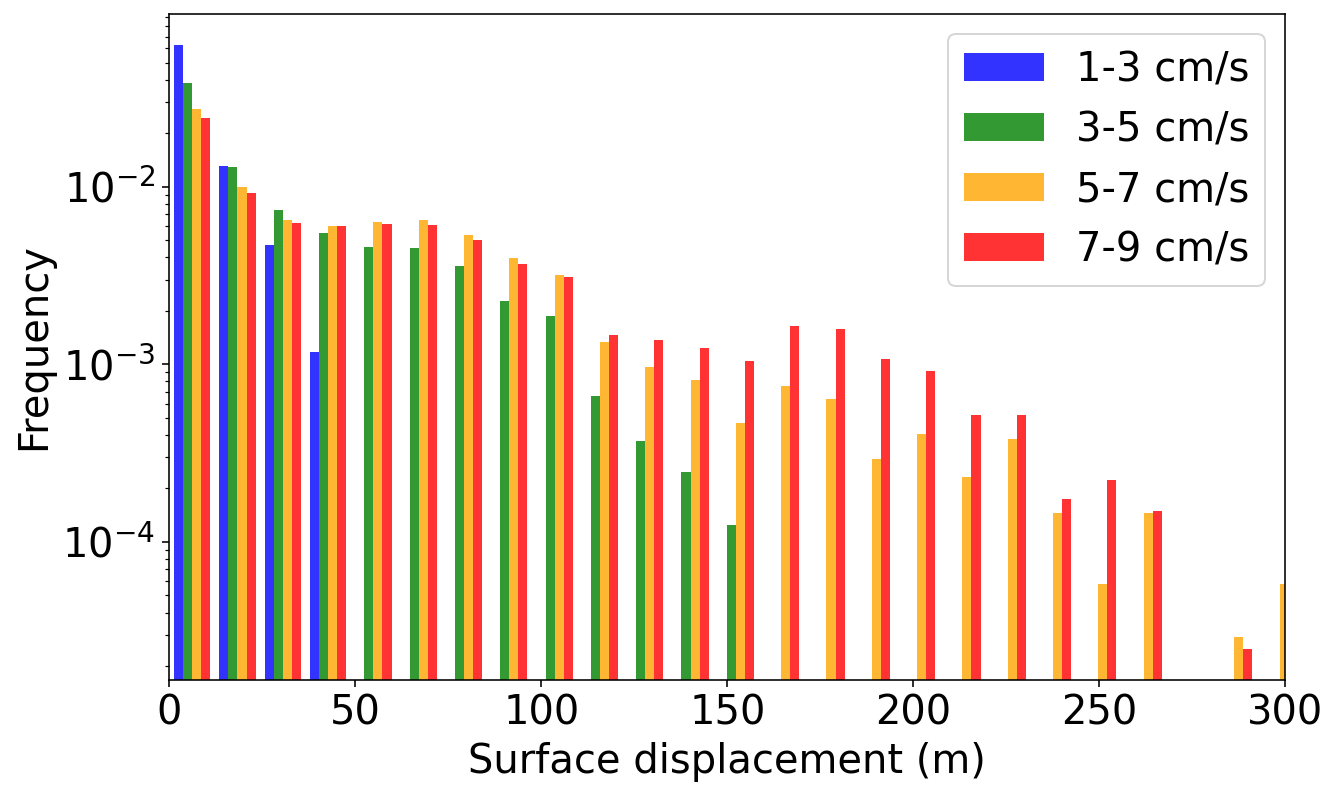}
  \caption{Distribution of the surface displacement of re-impacting particles on the DRDTM, for different ranges of initial ejection velocities. The surface displacement is measured from the first-contact point to the final resting position of each particle on the surface of Dimorphos. Colours indicate the initial ejection-velocity ranges: 1--3, 3--5, 5--7, and 7--9~cm\,s$^{-1}$. The corresponding mean displacements are 8.4, 27.7, 46.3, and 56.6~m, with standard deviations of 8.6, 30.6, 49.8, and 60.5~m, respectively.}
  \label{fig:histogram_displacement}
\end{figure}

\begin{figure}
  \centering
  \includegraphics[width=1.\linewidth]{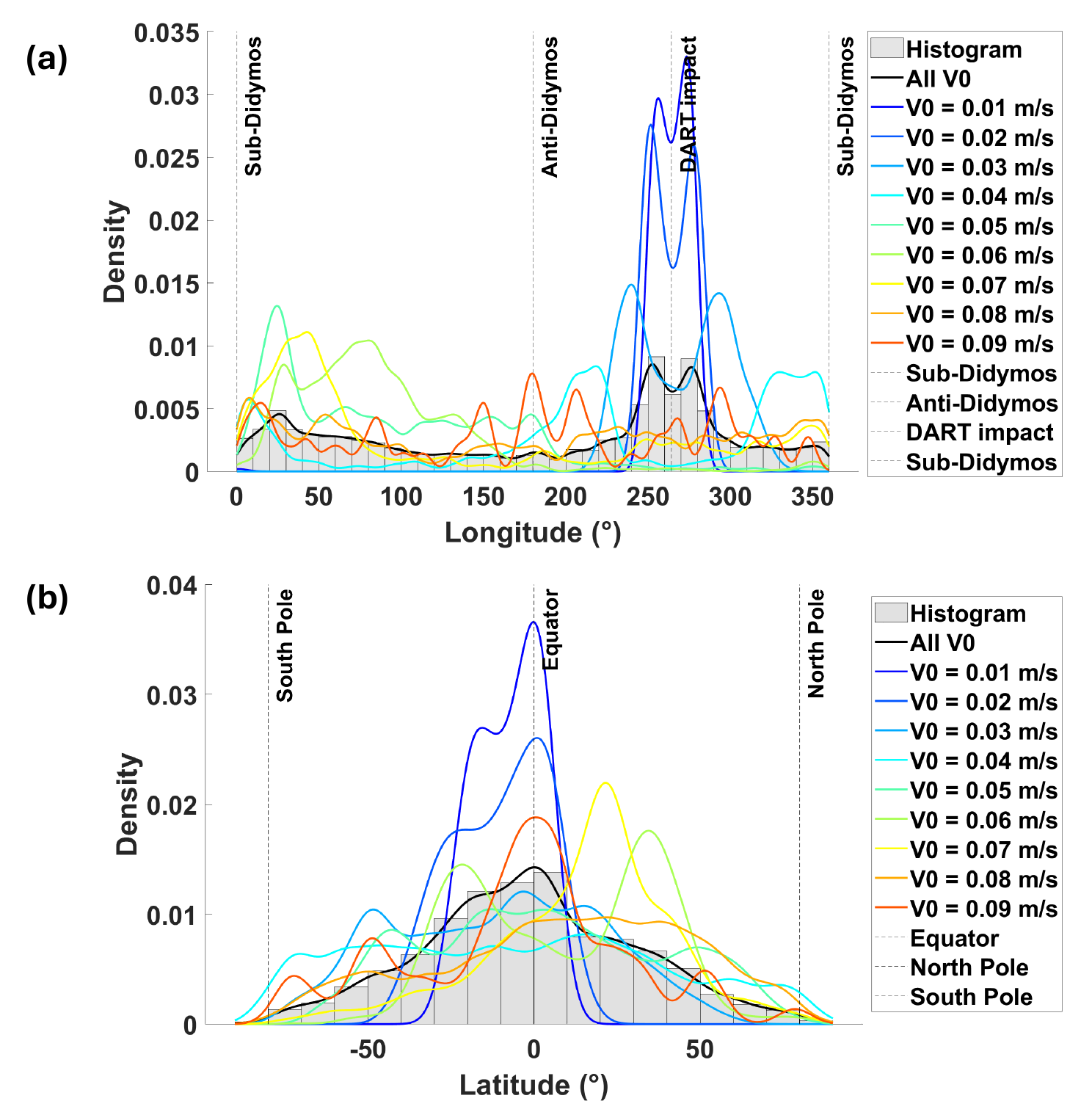}
  
  \caption{
    Longitude and latitude distributions of the final tracer positions for
    the smooth ellipsoidal model of Dimorphos with $R_c = 15$~m.
    Grey bars show the normalized histograms of all final positions, and
    black curves show KDEs computed from all end-points. Coloured curves
    show KDEs computed separately for each initial ejection velocity. The
    sharp peaks associated with the lowest velocities indicate that these
    tracers are more spatially concentrated around the DART impact region,
    whereas faster tracers produce broader distributions over longitude and
    latitude.
  }
  \label{fig:B2}
\end{figure}
% ------ end figs 15m-----------------

\section{Ejecta distribution on the DART/DRACO-based Dimorphos
         terrain model}
\label{sec:DART_DTM}

Because the smooth ellipsoidal model lacks topographic features, the
resulting surface trajectories of ejecta may be unrealistic. We therefore
adopt a more realistic shape model: the DART/DRACO-based Dimorphos terrain
model (\citealt{Daly_2024}; hereafter DRDTM) displayed in Fig. \ref{fig:DART_DTM}. Only the leading
hemisphere, which includes the DART impact site, was imaged and resolved,
while the opposite hemisphere remains smooth because of the lack of data
(Figs.~\ref{fig:figure_DART_DTM_impacts} and \ref{fig:DART_DTM}). We repeat the RAVEL simulations using identical
initial conditions but replacing the smooth shape by the DRDTM.

The orbital trajectories and first-contact distribution are essentially unchanged by the replacement of the smooth surface with the DRDTM (compare
Fig.~\ref{fig:figure_smooth_dimorphos}a and
Fig.~\ref{fig:figure_DART_DTM_impacts}a). The main differences emerge after
first contact, when 
tracers interact with the local surface topography (compare
Fig.~\ref{fig:figure_smooth_dimorphos}b,c and
Fig.~\ref{fig:figure_DART_DTM_impacts}b,c).

Inspection of the final deposition pattern reveals a clear effect of the
resolved surface topography. In the rough, well-resolved part of the DRDTM,
the final deposit forms elongated, topographically channelled depositional
streaks (Figs.~\ref{fig:figure_DART_DTM_impacts}b,c and
\ref{fig:3D_dart_model}). These structures are produced mainly by
low-velocity ejecta ($v<6$~cm\,s$^{-1}$), whereas tracers with
$v>7$~cm\,s$^{-1}$ contribute much less to them. The streaks develop as
returning ejecta are guided along valley-like features and locally trapped
or deflected by topographic highs
(Fig.~\ref{fig:3D_dart_model}).{ Although these structures are ray-like in morphology, ray formation is not
associated with a single mechanism. Previous studies have shown that
pre-existing surface relief can break the azimuthal symmetry of the ejecta
flow and generate preferential depositional directions during impact and
excavation \citep{Shuvalov_2012,Sabuwala_2018}. In our simulations, a
similar topographic control operates at a later stage: after re-accretion,
slow ejecta are guided, deflected, and trapped by the local relief during
surface transport after first contact. We therefore use the term
``ray-like'' as a morphological description, while referring to the
structures themselves as topographically channelled depositional streaks.} The rough topography does not appear to change the
dispersion extent for a given ejection velocity, as shown by the comparison
between Figs.~\ref{fig:figure_smooth_dimorphos}b and
\ref{fig:figure_DART_DTM_impacts}b. 
On the side antipodal to the DART
impact, the concentration of high-velocity ejecta is still observed, but it
is less affected by roughness because the trailing hemisphere is smooth
(by construction) in the DRDTM. {This velocity dependence is also apparent in the distance travelled across
the surface after first contact (Fig.~\ref{fig:histogram_displacement}).}

{For initial ejection velocities of 1--3, 3--5, 5--7, and
7--9~cm\,s$^{-1}$, the mean surface displacements are 8.4, 27.7, 46.3,
and 56.6~m, respectively, with corresponding standard deviations of
8.6, 30.6, 49.8, and 60.5~m. Thus, the slowest ejecta remain comparatively
close to their first-contact locations, whereas faster ejecta can travel
several tens of metres across the surface before reaching their final
resting positions.}

{An important limitation is that the DRDTM represents the pre-impact
surface of Dimorphos and therefore does not include the DART crater or the
impact-induced modification of the terrain within and immediately around
the crater. Crater excavation is expected to modify primarily the transient
crater and its immediate surroundings
\citep{Ormo_et_al_2022,Stickle2025}, whereas the depositional structures
obtained here extend beyond this region. Nevertheless, the actual
post-impact topography encountered by the re-accreting ejecta is presently
unknown, and constructing specific post-impact surface configurations would
require assumptions that are not currently constrained by observations. Consequently, we do not attempt to predict the exact locations,
orientations, or detailed morphology of individual depositional streaks.
Our prediction is more general: surface roughness can channel slow
re-accreted ejecta and generate elongated depositional structures. This
interpretation is further supported by the synthetic fully rough DTM
presented in Sect.~\ref{sec:fully_rough_DTM}, which has an independent topographic pattern but produces similar elongated structures.}

{We also verified that the emergence of these structures is robust to
different random realizations of the initial tracer population. Although
the detailed distribution of individual tracers and the morphology of the
streaks vary between realizations, elongated, topographically channelled
depositional structures are systematically recovered.}

In both the smooth and DRDTM cases, most of the mass accumulates around
the impact region and near the antipodal region
(Figs.~\ref{fig:figure_smooth_dimorphos}c and
\ref{fig:figure_DART_DTM_impacts}c). In the DRDTM case, however, the
near-impact mass concentration is additionally redistributed along the
preferential directions defined by the topographically channelled
depositional streaks (Fig.~\ref{fig:figure_DART_DTM_impacts}c). 

{The comparison between the smooth and rough cases therefore shows that,
within our simulations, surface topography alone can organize a broadly
distributed population of re-accreted ejecta into coherent preferential
depositional corridors after first contact.} 

%======================================================================

% ------------- Figs for 15m --------
\begin{figure*}[t]
  \centering

  \begin{minipage}[t]{0.49\linewidth}
    \centering
\includegraphics[width=\linewidth]{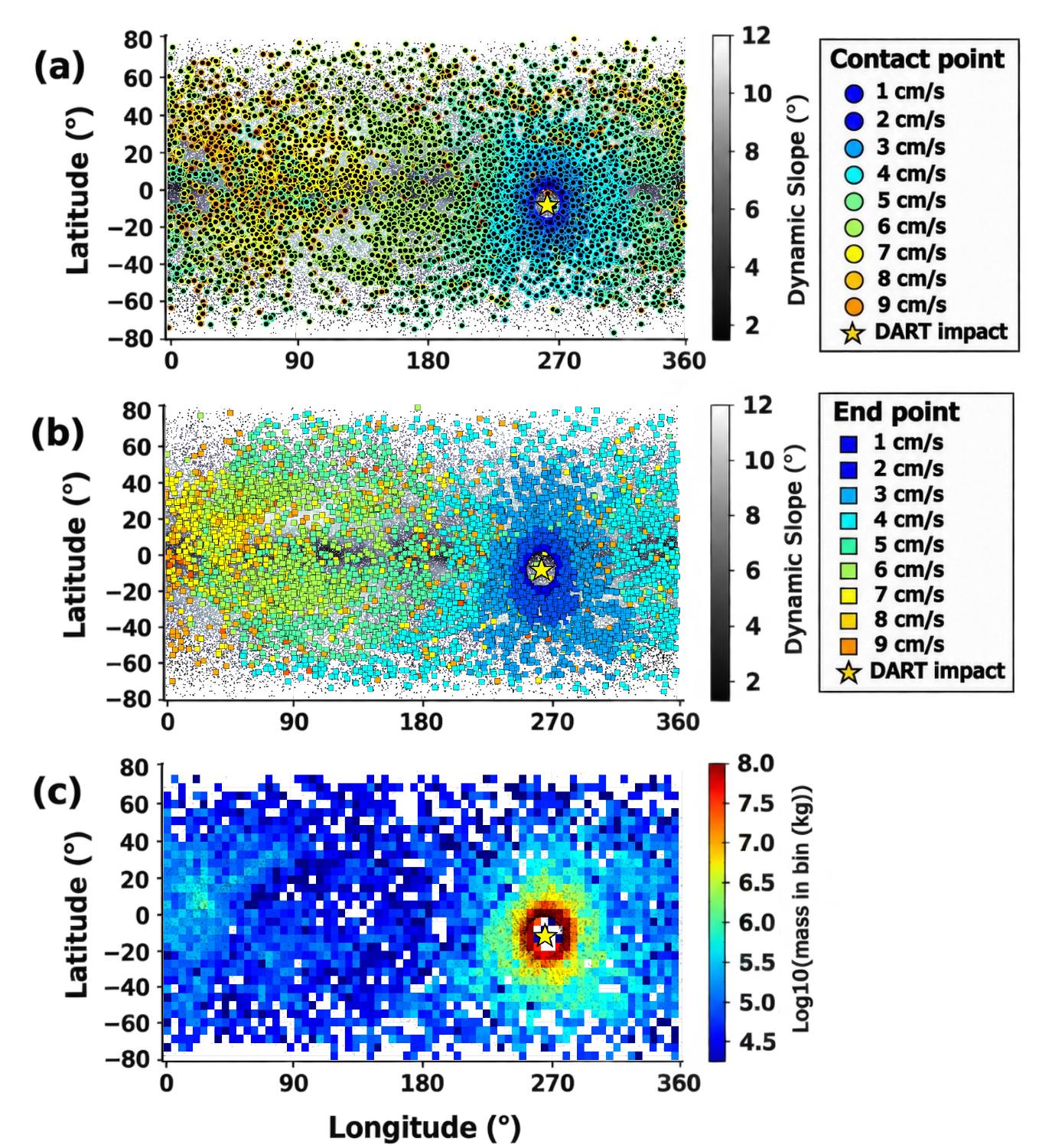}
  \caption{
    Same as Fig.~\ref{fig:figure_smooth_dimorphos}, but for an ejecta-source radius
    $R_c = 15$~m.
    (a) First-contact locations of the tracers on the smooth ellipsoidal
        model of Dimorphos.
    (b) Final resting positions after rebounds and frictional sliding.
    (c) Mass-weighted final distribution, expressed as $\log_{10}$ of the
        mass per bin ($5^\circ \times 5^\circ$).{ White regions correspond to
surface cells containing no deposited tracers at the end of the simulation and therefore represent zero deposited mass in our tracer
sampling.}
    Compared with the standard $R_c = 35$~m case, the lowest-velocity
    ejecta form a more compact near-impact deposit, while the large-scale
    velocity-dependent redistribution pattern remains essentially unchanged.
  }
  \label{fig:B1}
\end{minipage}
\hfill
\begin{minipage}[t]{0.48\linewidth}
    \centering
\includegraphics[width=\linewidth]{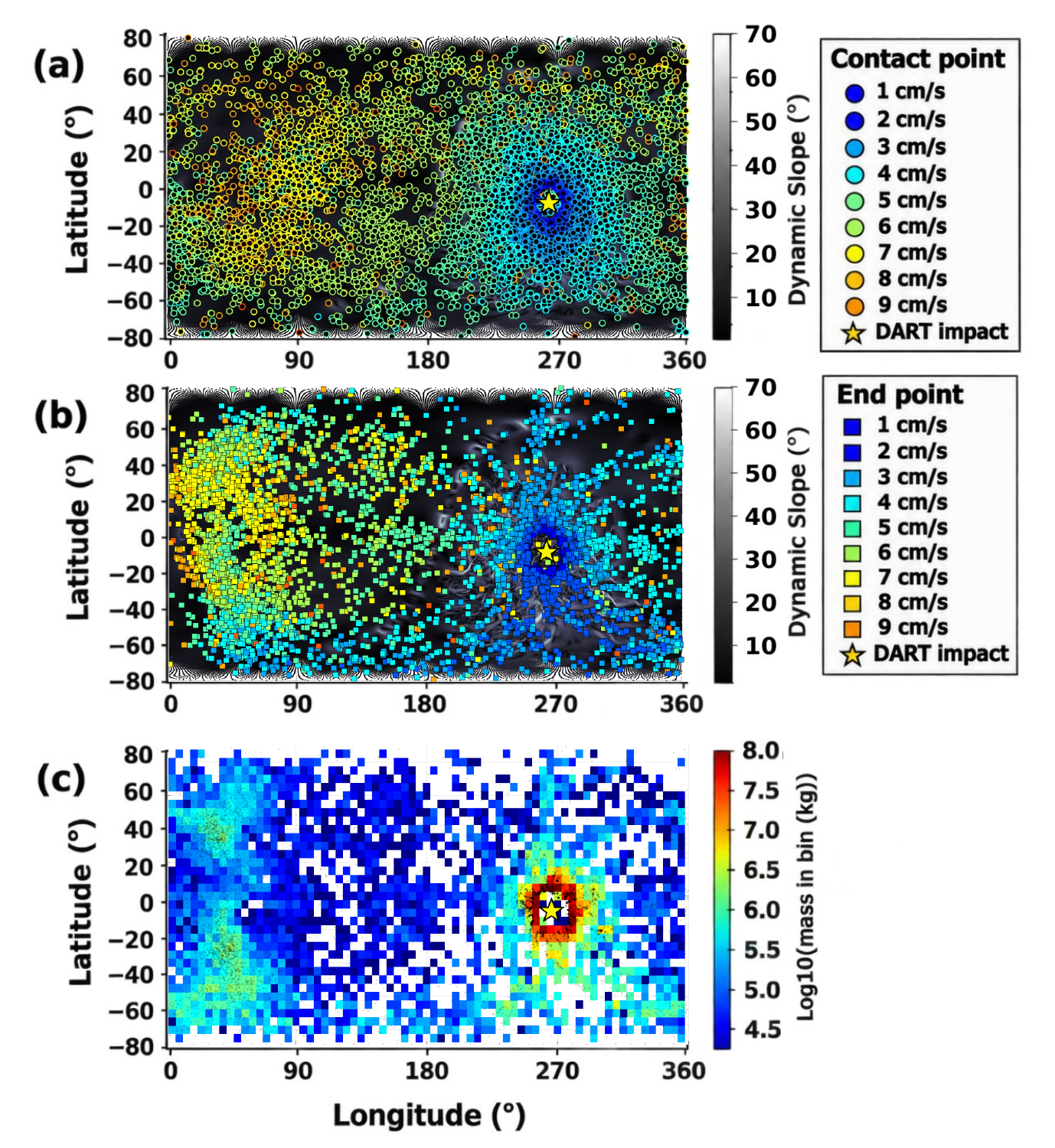}
  \caption{
    Distribution of ejecta on the DRDTM for an ejecta-source radius $R_c = 15$~m.
    (a) First-contact locations of the tracers on the surface.
    (b) Final resting positions after rebounds and frictional sliding.
    (c) Mass-weighted final distribution, expressed as $\log_{10}$ of the
        mass per bin ($5^\circ \times 5^\circ$).{ White regions correspond to
surface cells containing no deposited tracers at the end of the simulation and therefore represent zero deposited mass in our tracer
sampling.}
  }
  \label{fig:15M_DART-DTM}
\end{minipage}

\end{figure*}

\begin{figure}
  \centering
  \includegraphics[width=0.935
  \linewidth]{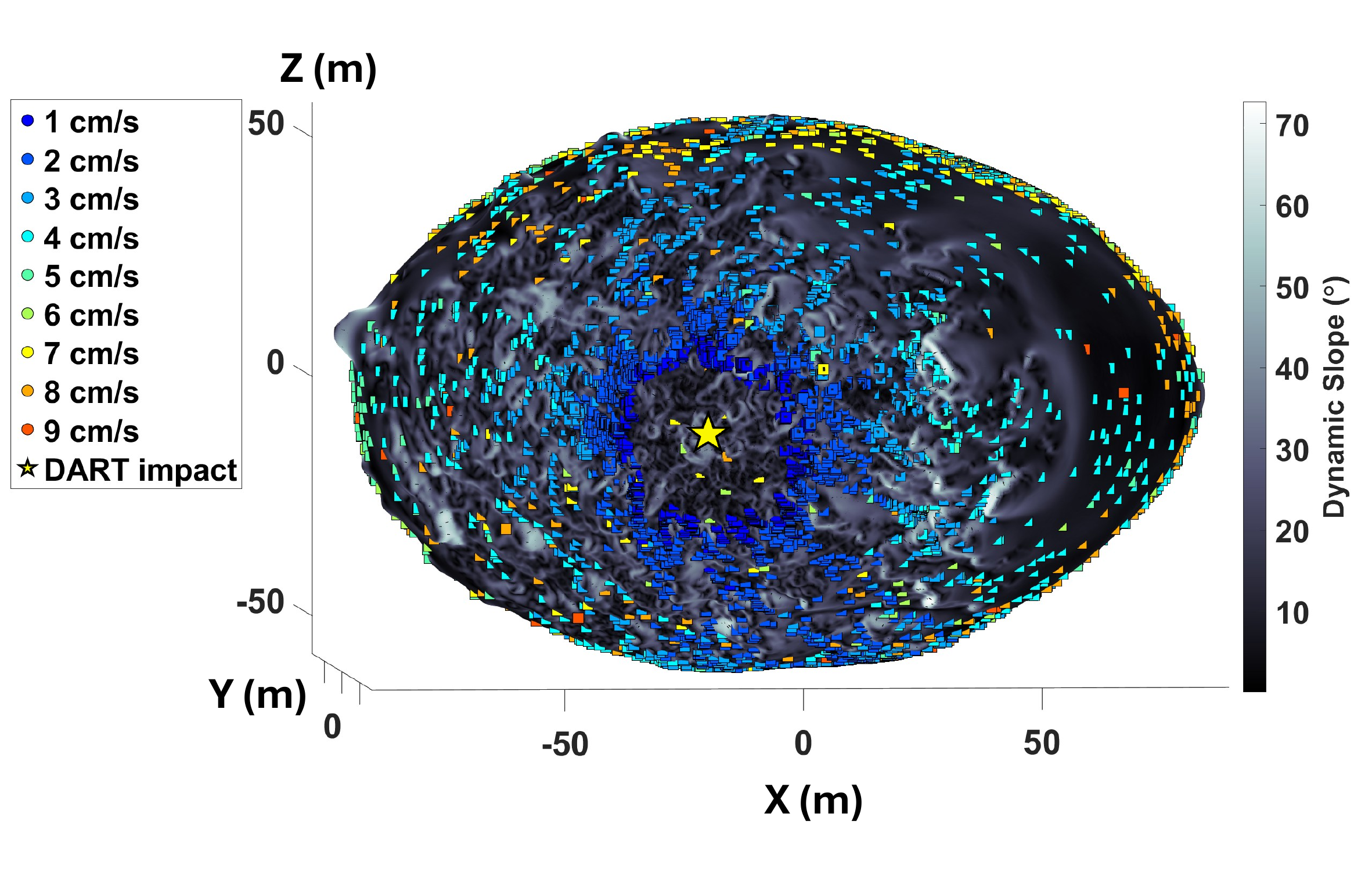}
  \caption{
    Three-dimensional view of the final ejecta distribution on the DRDTM
    for $R_c = 15$~m. Colours indicate the initial ejection velocity, and
    the grey-scale surface shows the local dynamic slope. The slowest
    ejecta are channelled by the rough topography around the DART impact
    site, producing ray-like deposits that can persist into the smooth,
    non-imaged part of the DTM once the material has been guided into
    preferential pathways.
  }
  \label{fig:15m_3D}
\end{figure}
%======================================================================

\section{Sensitivity to the spatial extent of the ejecta source: the $R_c=15$ m case}
\label{sec:radius_15m}

{Given that the DART crater size remains uncertain, we repeated the
simulations using a smaller ejecta source radius, $R_c = 15$~m, to assess
the influence of this parameter on our results. All other parameters were
kept identical to the standard case, including the ejection-velocity range
and mass--velocity distribution based on \citet{Stickle2025}, the
ejection-angle range observationally motivated by
\citet{Hirabayashi_2025}, the friction angle, the coefficient of
restitution, and the integration time. Since the initial distance of the
tracers from the source centre scales linearly with $R_c$ in
Eq.~(\ref{eq:Housen_Holsapple_law}), this test isolates the effect of the
spatial extent of the initial ejecta source while leaving the prescribed
ejecta properties and the subsequent orbital and surface dynamics
unchanged.}

The resulting distributions are shown in
Figs.~ \ref{fig:B2},~\ref{fig:B1}, \ref{fig:15M_DART-DTM}, and
\ref{fig:15m_3D}. {Although differences with the standard $R_c = 35$~m case are observed close to the ejection zone, where the lowest-velocity ejecta
are concentrated, the same large-scale behaviour is recovered. The final ejecta distribution remains strongly asymmetric and velocity-dependent. Low-velocity tracers remain preferentially concentrated around the DART impact region, whereas faster tracers are transported over much larger distances and preferentially populate distal and antipodal regions. This indicates that, within the prescribed low-velocity ejecta population, the
large-scale dynamical sorting is controlled primarily by ejection velocity
and the binary dynamical environment, and is only weakly affected by the
spatial extent of the source region considered here.}

The main difference with respect to the standard $R_c=35$~m case is the
stronger spatial concentration of the slowest ejecta around the impact
site. In the first-contact and final end-point maps, tracers with the
lowest ejection velocities form a more compact cluster around the DART
impact longitude and latitude. This effect is particularly clear in the
mass-weighted map, where the highest surface mass densities remain confined
to a compact ejecta blanket surrounding the impact region. Because the
adopted mass--velocity distribution assigns most of the ejecta mass to the
lowest velocities, reducing $R_c$ primarily affects the compactness of this
high-mass near-impact deposit.

We also repeated the $R_c=15$~m case using the DRDTM to assess the combined
effect of a smaller {source region} and the resolved rough topography around
the DART impact site (Fig.~\ref{fig:15M_DART-DTM}). As in the smooth
$R_c=15$~m case, reducing the spatial extent of the source produces a more
compact concentration of the lowest-velocity ejecta around the impact
region. However, interaction with the DRDTM breaks the azimuthal symmetry
of this compact deposit and organizes part of the slow ejecta into
{topographically channelled depositional streaks}. Compared with the standard
$R_c=35$~m DRDTM case, these structures remain present and extend over
comparable distances, reflecting the preferential pathways imposed by the
same underlying terrain. The distal accumulation of faster ejecta in the
antipodal and trailing hemispheres is also recovered, indicating that the
large-scale velocity-dependent redistribution remains controlled primarily
by the binary dynamical environment, while the rough topography mainly
modifies the local morphology of the deposits after re-impact.

The $R_c = 15$~m case therefore emphasizes the compactness of the
lowest-velocity ejecta blanket, but it does not change the main conclusions
obtained for the standard $R_c = 35$~m case. The large-scale distribution
of ejecta remains highly asymmetric and velocity-sorted, and is substantially
modified by {surface transport after first contact}. The DRDTM calculation further shows
that local topography can organize slow ejecta into topographically channelled depositional streaks even when the spatial
extent of the initial source region is reduced. Faster ejecta still produce broad longitudinal and latitudinal
distributions, whereas the slowest ejecta, {which dominate the mass under our fiducial weighting}, remain
concentrated close to the DART impact site.

%======================================================================
\section{Effect of global roughness: a synthetic fully rough Dimorphos DTM}
\label{sec:fully_rough_DTM}

The DRDTM provides a realistic description of the imaged hemisphere of
Dimorphos, including the region surrounding the DART impact site. However,
the opposite hemisphere was not resolved by DRACO and is therefore
intentionally smooth in the available DTM. This limitation prevents us from
assessing how roughness would affect ejecta transported beyond the imaged
region, in particular the distal deposits produced by the faster tracers.
To explore this effect, we constructed a synthetic fully rough Dimorphos
DTM, in which topographic roughness is distributed over the entire surface.
This model is not intended to reproduce the actual topography of the
non-imaged hemisphere, but to provide a sensitivity test of the role of
global roughness in shaping the final ejecta distribution.

\begin{figure}
  \centering
  \includegraphics[width=0.9\linewidth]{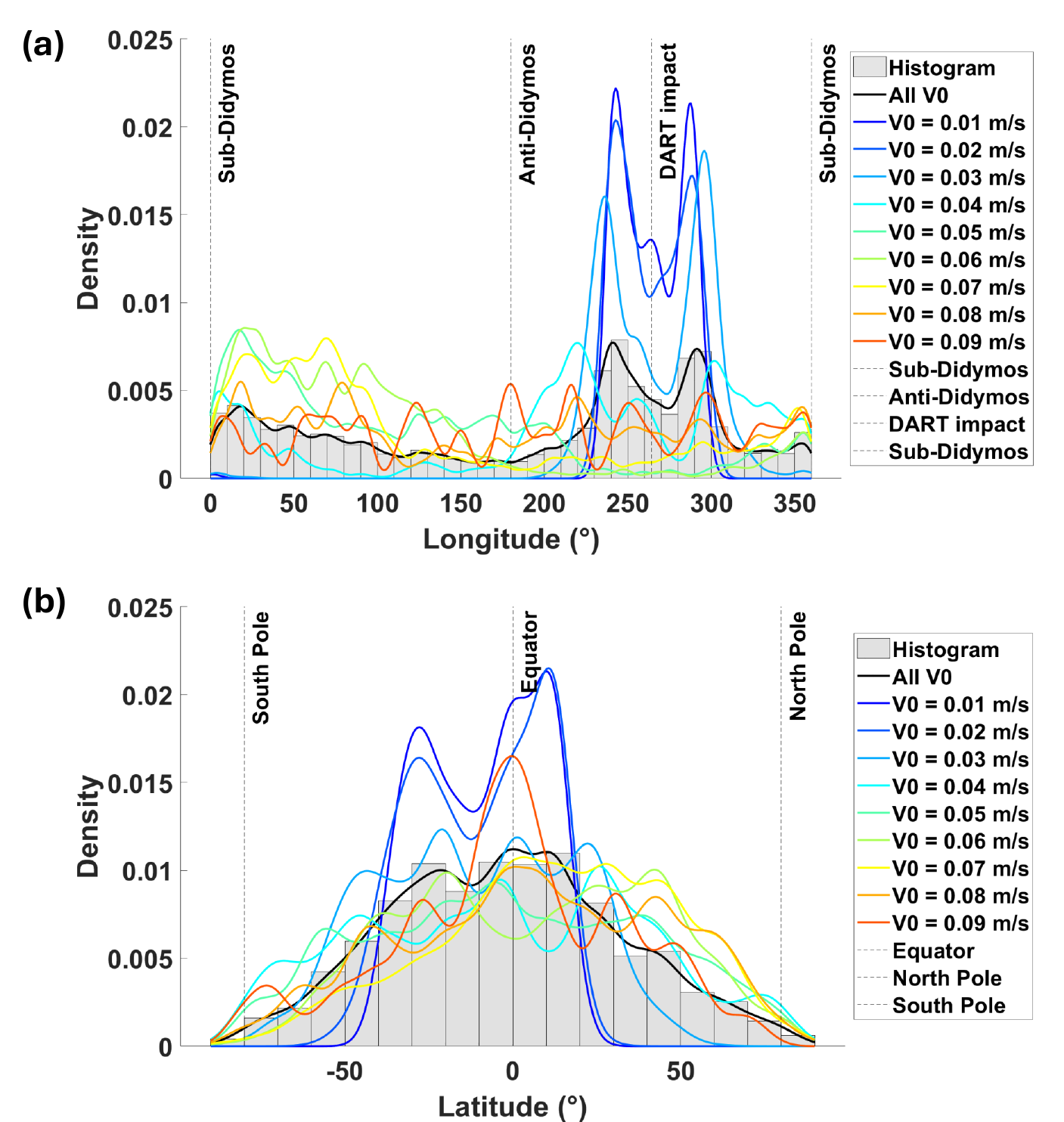}
  \caption{
    Longitude and latitude distributions of the final tracer positions for
    the synthetic fully rough Dimorphos DTM ($R_c = 35$~m). Grey bars show
    the normalized histograms of all final positions, and black curves show
    KDEs computed from all end-points. Coloured curves show KDEs computed
    separately for each initial ejection velocity. The lowest-velocity
    ejecta remain concentrated around the DART impact region, but the fully
    rough surface allows the associated ray-like structures to extend {farther from the impact region}. Higher-velocity ejecta still populate broad
    distal regions, although their final distribution is more dispersed
    than in the smooth ellipsoidal case.
  }
  \label{fig:C2}
\end{figure}

\begin{figure}
  \centering
  \includegraphics[width=1\linewidth]{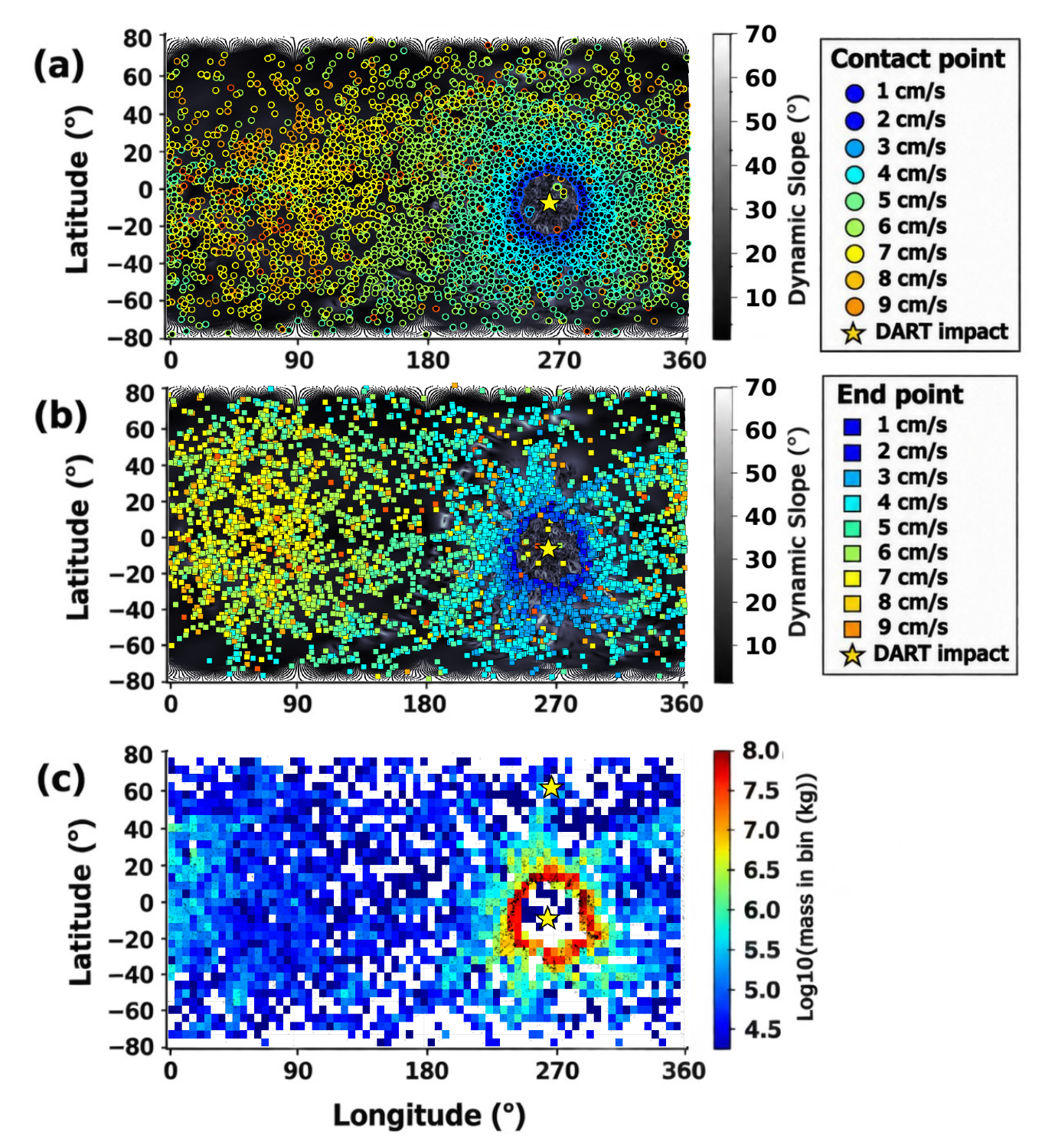}
\caption{
  Distribution of ejecta on the synthetic fully rough Dimorphos DTM for
  $R_c = 35$~m and ejection angles 23.5$^{\circ}$--42.5$^{\circ}$. Initial velocities and launch positions are related through Eq. (\ref{eq:Housen_Holsapple_law}).
  (a) First-contact locations of the tracers on the surface.
  (b) Final resting positions.
  (c) Mass-weighted final distribution.
  The large-scale velocity-dependent redistribution pattern is preserved,
  while the globally rough surface produces elongated, topographically
  guided depositional structures.
}
  \label{fig:C1}
  \end{figure}

The results obtained with this synthetic fully rough DTM for $R_c = 35$~m are shown in Figs.~\ref{fig:C2} and \ref{fig:C1}.
The first-contact distribution remains broadly similar to that obtained in
the smooth ellipsoidal case. Low-velocity tracers re-impact close to the
DART impact site, whereas faster tracers are transported over larger
distances and can reach the antipodal and trailing hemispheres. This
confirms that the primary velocity sorting is mainly controlled by the
initial ejection velocity and by the binary dynamical environment, rather
than by the details of the surface topography.

The main differences appear after the first contact with the surface.
Compared with the smooth ellipsoid, the final end-point distribution becomes
more structured and spatially heterogeneous. The rough surface locally
guides, deflects, traps, or disperses tracers during rebounds and
frictional sliding. As a result, the final distribution no longer reflects
only the large-scale orbital transport, but also the interaction between
ejecta and local topographic features. This confirms the conclusion reached
with the DRDTM: surface roughness does not modify the preceding orbital
trajectories, but it can strongly affect the final depositional pattern
once tracers interact with the surface.

%{Importantly, elongated depositional streaks also emerge on this independently generated rough surface despite its different topographic pattern, whereas they are absent from the smooth ellipsoidal model. Their detailed locations and orientations depend on the underlying terrain, but their emergence on both rough surfaces supports topographic channelling as a robust qualitative outcome of the simulations.}

 {Importantly, the final ejecta distribution on this independently
generated rough surface is highly heterogeneous and spatially
clustered, with elongated concentrations of particles (streaks)
and regions strongly depleted of ejecta.
These structures result from the interaction between particle
trajectories and local surface relief.
Topographic channelling redirects particles along preferential
pathways, producing elongated concentrations, whereas
deflections by prominent topographic features can prevent
deposition in adjacent regions, producing topographic shadowing.
These two mechanisms are illustrated in
Appendix~\ref{annexe_topographic_effects}, where individual
trajectories reveal coherent accumulations and deflections associated with
specific topographic features.
The absence of comparable structures on the smooth ellipsoid
supports their physical origin and indicates that their
detailed morphology is controlled by the local topography.}
{In the fully synthetic rough case, roughness is present over the entire
surface, and elongated depositional structures can extend over large
distances, up to $\sim270^\circ$ around Dimorphos in the simulation shown
here. The calculation uses the same three ejection angles
($23.5^\circ$, $33^\circ$, and $42.5^\circ$) and the same remaining model
parameters as the DRDTM simulations. It therefore provides a direct
sensitivity test of the effect of extending surface roughness beyond the
DRACO-imaged hemisphere. The $\sim270^\circ$ extent should not be
interpreted as a prediction of the exact spatial extent of such structures
on Dimorphos, since the synthetic topography does not reproduce the actual
terrain of the non-imaged hemisphere. Rather, this experiment shows that,
when roughness is present over the entire surface, topographic channelling
of re-accreted ejecta can persist far from the impact region. Since the slowest ejecta dominate the mass budget, the mass-weighted map provides a quantitative estimate of the deposited mass associated with these topographically guided ray-like structures around the DART impact site.}

The fully rough DTM also modifies the distal deposits associated with faster
ejecta. In the smooth ellipsoidal case, high-velocity tracers preferentially
accumulate in the antipodal region, particularly on the trailing hemisphere.
This large-scale accumulation is still present in the fully rough
simulation, supporting its primarily dynamical origin. However, because the faster
tracers now re-impact and move over a rough distal terrain, their final
resting positions become more scattered. Global roughness does not suppress
the dynamical focusing of fast ejecta towards the opposite hemisphere, but it
redistributes this material locally after re-impact through rebounds,
frictional sliding, and topographic deflection.

%-------- Figs fully rough -----
%-------------------------------------------------

%--- end figs fully rough ---------
%----- Figs. random ---------------
\begin{figure*}[t]
  \centering

  \begin{minipage}[t]{0.49\linewidth}
    \centering
    \includegraphics[width=\linewidth]{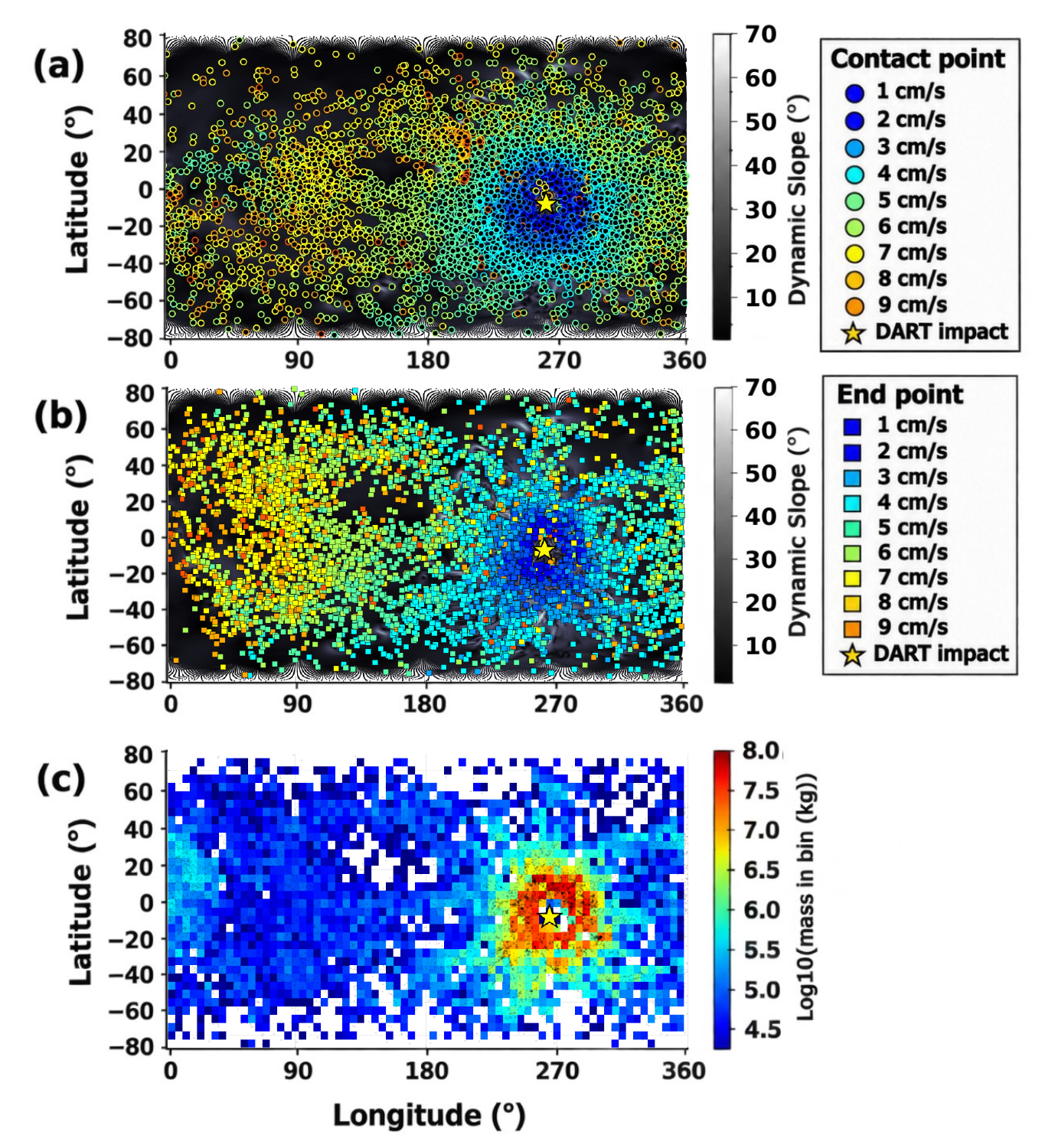}
  \caption{
    {Distribution of ejecta on the DART/DRACO-based DTM after
removing the correlation between launch position and ejection velocity.
Initial positions are randomly distributed between $R_c/3=11.6$~m and
$R_c=35$~m from the DART impact point, while initial velocities are drawn
independently from a uniform distribution between 1 and 9~cm\,s$^{-1}$.
(a) First-contact locations.
(b) Final resting positions.
(c) Mass-weighted final distribution, expressed as $\log_{10}$ of the
mass per bin ($5^\circ\times5^\circ$).
White regions correspond to surface cells containing no deposited tracers
at the end of the simulation and therefore represent zero deposited mass
in our tracer sampling.
This case should be compared with Fig.~\ref{fig:figure_DART_DTM_impacts},
in which launch position and velocity are related through
Eq.~(\ref{eq:Housen_Holsapple_law}).}}
  \label{fig:figure_DART_DTM_RANDOM_impacts}
\end{minipage}
  \hfill%
  \begin{minipage}[t]{0.48\linewidth}
    \centering
    \includegraphics[width=\linewidth]{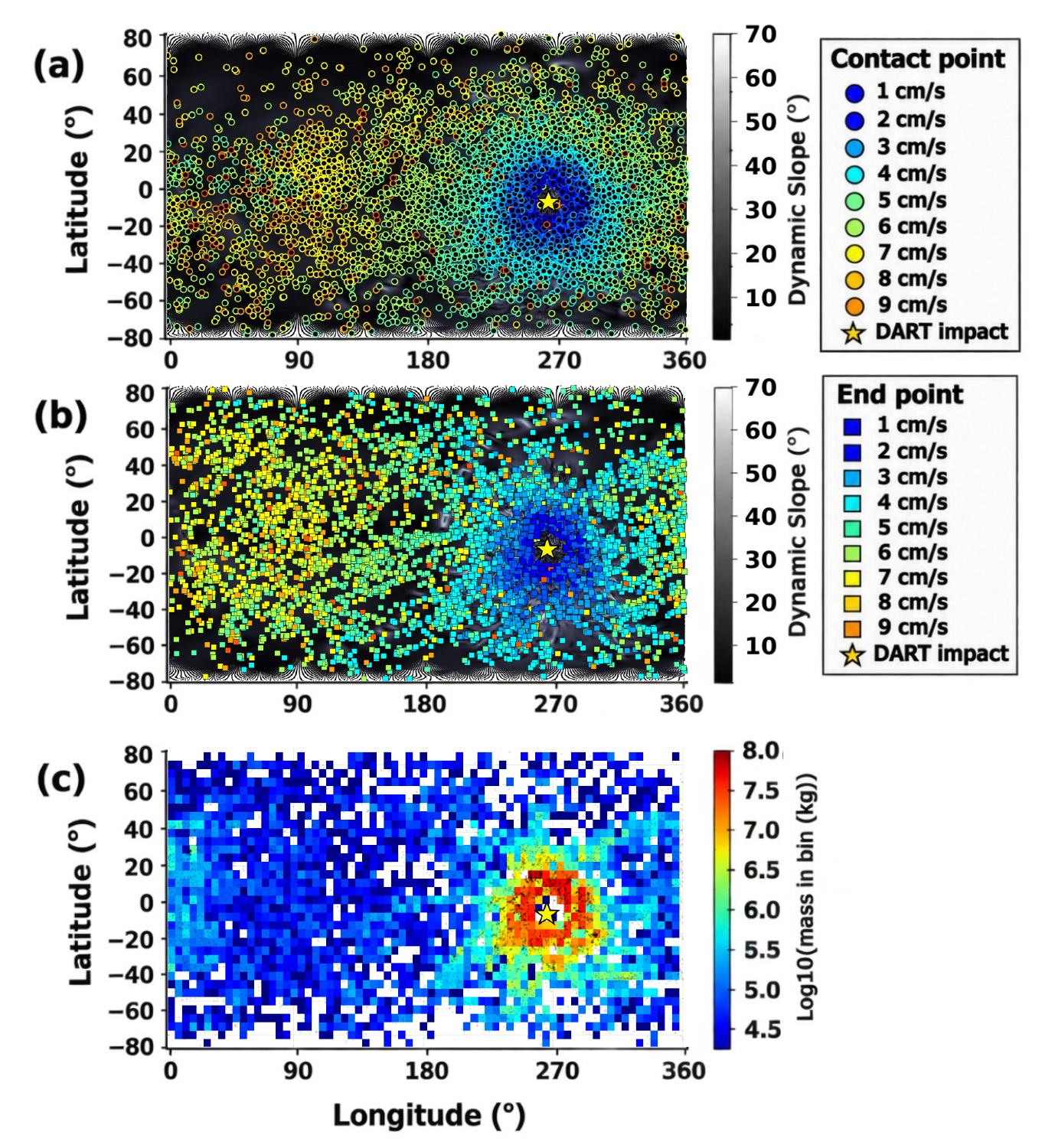}
\caption{
  Distribution of ejecta on the synthetic fully rough Dimorphos DTM for
  $R_c = 35$~m and ejection angles 23.5$^{\circ}$--42.5$^{\circ}$. Initial velocities and launch positions are randomly distributed within an annulus extending from $R_c/3$ to $R_c$.
  (a) First-contact locations of the tracers on the surface.
  (b) Final resting positions.
  (c) Mass-weighted final distribution.
  The large-scale velocity-dependent redistribution pattern is preserved,
  while the globally rough surface produces elongated, topographically
  guided depositional structures.
}
  \label{fig:rough_random}
  \end{minipage}

\end{figure*}

%-------------------------------
%------ end figs random ----

\section{Sensitivity to the launch position-velocity relation} \label{sec:random_initial_conditions}

{In the simulations presented above, the initial launch position and velocity are related through Eq.~(\ref{eq:Housen_Holsapple_law}). Because the relation between launch position and ejection velocity on
Dimorphos remains uncertain, we tested whether our main result, i.e. the emergence of ray-like depositional structures around the DART crater,  depends on this assumed position-velocity relation. We therefore performed an extreme sensitivity test in which the relation between position and velocity given in Eq. (\ref{eq:Housen_Holsapple_law}) was completely removed. As in the standard simulations, tracer launch positions were randomly distributed within an annulus extending from $R_c/3$ to $R_c$, corresponding to
$11.6$--$35$~m from the crater centre, using the same annulus dimensions as in Sect.~\ref{sec:results_smooth}. However, instead of assigning velocities according to Eq.~(\ref{eq:Housen_Holsapple_law}), each tracer was assigned an initial velocity drawn independently from a uniform distribution between 1 and 9~cm~s$^{-1}$.}

{The resulting final ejecta distributions are shown in
Figs.~\ref{fig:figure_DART_DTM_RANDOM_impacts} and
\ref{fig:rough_random} for the DRDTM and the synthetic fully rough DTM,
respectively. Despite the complete removal of the launch
position--velocity correlation, the overall depositional patterns remain
similar to those obtained using the Housen--Holsapple relation
(Figs.~\ref{fig:figure_DART_DTM_impacts} and \ref{fig:C1}),
including both the near-impact concentration and the topographically
guided ray-like structures. }

{The main differences are confined to the launch region, particularly near
its inner boundary at $R_c/3$ ($11.6$~m here). In the randomized case, low-velocity tracers ($<2$~cm~s$^{-1}$) can be launched from the inner region and therefore remain or re-impact nearby. In contrast, the Housen--Holsapple prescription preferentially associates the lowest ejection velocities with launch positions near the outer region of the ejecta annulus ($R_c=35$~m). }

{We therefore conclude that the assumed launch position--velocity relation has only a secondary influence on the final depositional pattern and primarily affects material within or close to the initial ejecta annulus. In particular, the ray-like structures extending beyond the launch region persist when the position--velocity correlation is completely removed. Their formation is therefore not an artefact of the adopted Housen--Holsapple scaling law, but results primarily from the interaction of returning ejecta with the local topography. The detailed distribution
within the source region depends on the assumed $v(r)$ relation, but the
large-scale depositional pattern and the ray-like structures do not. 
}

\section{Summary and conclusions}
\label{sec:conclusions}

{We investigated the first $\sim22$~h evolution of low-velocity DART
impact ejecta on Dimorphos with RAVEL (Sect.~\ref{sec:method}), coupling
orbital motion in the Didymos--Dimorphos system with surface transport
after first contact. The first surface contacts are strongly velocity
sorted, and more than 95\% of the simulated tracers re-accrete within
$\sim5$~h. Slow ejecta ($v\lesssim4$~cm\,s$^{-1}$) rapidly return near
the DART impact region and remain preferentially concentrated there,
whereas faster tracers can undergo partial orbits around Dimorphos or
temporary circulation around Didymos before returning. Using our reference mass--velocity distribution given in {Eq. (\ref{eq:mass_velocity})  \citep{Stickle2025}}, the slowest ejecta dominate the deposited mass, while the faster ejecta represent a much smaller mass fraction and
preferentially populate distal, antipodal, and trailing regions. The
depletion obtained near longitudes $\pm90^\circ$ from the DART impact
site is consistent with previous studies of the long-term evolution of
DART ejecta \citep{Langner_2024,Langner_2025}.
}

{An important result of this study is that the final ejecta distribution
cannot be understood from orbital dynamics alone. After re-impact,
particles can travel substantial distances across the surface before
reaching their final resting positions. In the DRDTM simulations, the
mean surface displacements increase from 8.4~m for ejecta with initial
velocities of 1--3~cm\,s$^{-1}$ to 56.6~m for the 7--9~cm\,s$^{-1}$
population. Surface transport after first contact is therefore not a
minor correction to the ballistic evolution, but an important component
of the redistribution of low-velocity ejecta on Dimorphos.}

 {Using the DART/DRACO-based rough terrain model of Dimorphos
(DRDTM), we find that the interaction between re-accreted
ejecta and local topography strongly structures the final
deposition pattern, producing elongated concentrations of
particles (streaks) and regions depleted of ejecta.
These features arise from topographic channelling, deflection,
and shadowing during surface transport after first contact,
as illustrated by individual trajectories in
Appendix~\ref{annexe_topographic_effects}.}
Similar structures emerge on the independently generated
fully rough synthetic surface, whereas they are absent
from the smooth ellipsoidal model.
This supports a physical origin for the observed spatial
clustering and depletion patterns, although their detailed
locations and morphology depend on the underlying terrain.

{The robustness of this result is further supported by the sensitivity
tests. The ray-like structures persist when the assumed launch
position--velocity relation $v(r)$ is completely removed, showing that
their formation is not an artefact of the adopted Housen--Holsapple
scaling of {Eq. (\ref{eq:Housen_Holsapple_law})}. Changing the source radius from $R_c=35$~m to $R_c=15$~m mainly
affects the compactness of the lowest-velocity deposit around the impact
region, while preserving the large-scale velocity sorting and the
formation of topographically guided structures { (i.e. ray-like streaks)}. Different random
realizations likewise modify the detailed tracer distribution without
suppressing the emergence of the ray-like pattern.
}

{The DRDTM represents the pre-impact surface and does not include the DART
crater or the topographic modification produced during crater excavation.
The actual post-impact terrain will only be known once \textit{Hera}
observes Dimorphos. However, crater excavation is expected to modify
primarily the crater and its immediate surroundings
\citep{Ormo_et_al_2022,Stickle2025}, whereas the structures obtained here
extend beyond this region. Moreover, their formation on an independent
synthetic rough surface shows that the mechanism does not rely on the
specific pre-impact topography of the DRDTM. We therefore do not predict
the exact location, orientation, or morphology of individual streaks on
the post-DART surface; rather, our result is that rough topography can
organize slow re-accreted ejecta into elongated depositional structures.
}

{Ray formation is not associated with a unique mechanism. Previous studies
have shown that pre-existing surface relief can generate preferential
ejecta directions during impact and excavation
\citep{Shuvalov_2012,Sabuwala_2018}, while target heterogeneity,
particle-size distributions, or projectile geometry may also produce
ray-like or filamentary patterns
\citep{Kadono_et_al_2019,Kadono_et_al_2022,Ormo_et_al_2022,
Pacheco-Vazquez_2019}. RAVEL does not simulate the impact or excavation
process and therefore does not address these impact-stage mechanisms.
Instead, our calculations demonstrate that ray-like depositional
structures can also arise exclusively during the subsequent
re-accretion and surface-transport stages: slow ejecta returning to the
surface are channelled by local relief after first contact. To our
knowledge, this is the first numerical demonstration of topographically
controlled ray-like depositional structures forming through re-accretion
and subsequent surface transport on a small body.
}

{\textit{Hera} will provide a direct opportunity to test these predictions.
\textit{HyperScout-H} \citep{Popescu_2025_Hyperscout} may identify
spectral units and freshly exposed or redistributed material associated
with the DART ejecta blanket, while the Asteroid Framing Cameras
(AFCs; \citealt{Vincent2026AFC}) will map albedo patterns and
geomorphological features at high spatial resolution. The spatial
distribution of the ejecta, the presence of ray-like depositional
structures and their relationship with local relief, and the extent of
antipodal accumulation can therefore provide observational constraints
on the importance of surface mobility after re-impact. Comparison with
first-order dynamical models such as RAVEL may then guide future,
more detailed modelling of regolith mechanics under microgravity.}

%\clearpage

\section*{Acknowledgement}
This research is supported by the French ANR project Roche, number ANR-23-CE49-0012, and the French Space Agency (CNES). {We thank Frédéric Durillon (www.animea.com) for his help with the development of the synthetic fully rough Dimorphos DTM. We thank the anonymous referee for the constructive comments, which improved the quality of the paper, and the Editor  for his careful handling of the manuscript.} 

\bibliography{bibliography}

%======================================================================

\appendix
\nolinenumbers

\section{Validation of surface curvature treatment in RAVEL}
\label{annex_section_deteachement_sphere}
\begin{figure}

  \centering

  \includegraphics[width=1.1\linewidth]{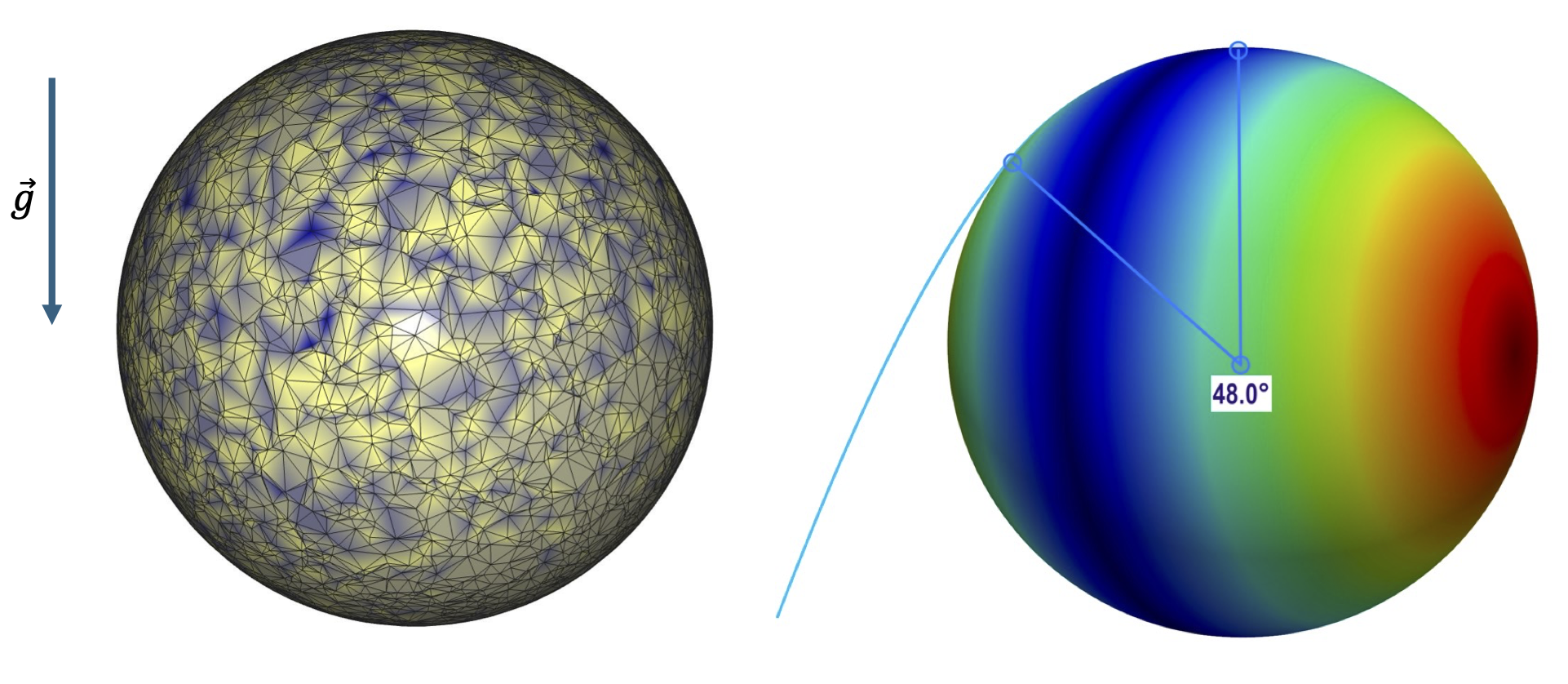}

  \caption{Left: 3D visualization of the polyhedral sphere used in the contact test, with $10^4$ nodes (for illustrative purposes). Colours indicate the distance from the sphere centre, and the arrow indicates the direction of gravity. Right: 3D visualization of one particle trajectory (blue line), released from rest  the north pole of a sphere with 10$^6$ nodes, illustrating the measurement of the detachment angle. Colours indicate the local angle between the gravity vector $\mathbf{g}$ and the surface normal.}

  \label{Figure_annex_3D_deteachment}

\end{figure}

 {The numerical treatment of surface contact in RAVEL is described in detail in \cite{Herreros_Charnoz_Phobos_2026}. In RAVEL, the surface is represented by a polyhedron composed of triangular facets (Figure \ref{Figure_annex_3D_deteachment}, left). Each facet is locally planar, and the normal reaction is computed using its local normal vector. At each time step, the velocity component directed into the surface is removed \citep{Herreros_Charnoz_Phobos_2026}, while particles remain free to detach and follow ballistic trajectories. Thus, surface geometry is accounted for through variations in local normals and successive contact corrections, without introducing an explicit curvature term.}

 {To validate this approach, we performed a test involving particles sliding without friction on a triangulated sphere under uniform vertical gravity (Figure \ref{Figure_annex_3D_deteachment}, left). Each simulation consists of 300 particles initially released within 1$^\circ$ of the North Pole of a sphere of $R$ = 1000 m radius in the Earth gravity field with $g$ = 9.81 m s$^{-2}$ and timestep $\Delta t$ = 0.1 s. From classical Newtonian mechanics, the analytical detachment angle for a particle released from rest at the top of a smooth sphere is $arccos(2/3)=48.19^\circ$ \citep{Prior2007_sliding_sphere}. Our simulations closely reproduce this behaviour (Figure \ref{Figure_annex_3D_deteachment}, right).}

\begin{figure}

  \centering

  \includegraphics[width=0.9\linewidth]{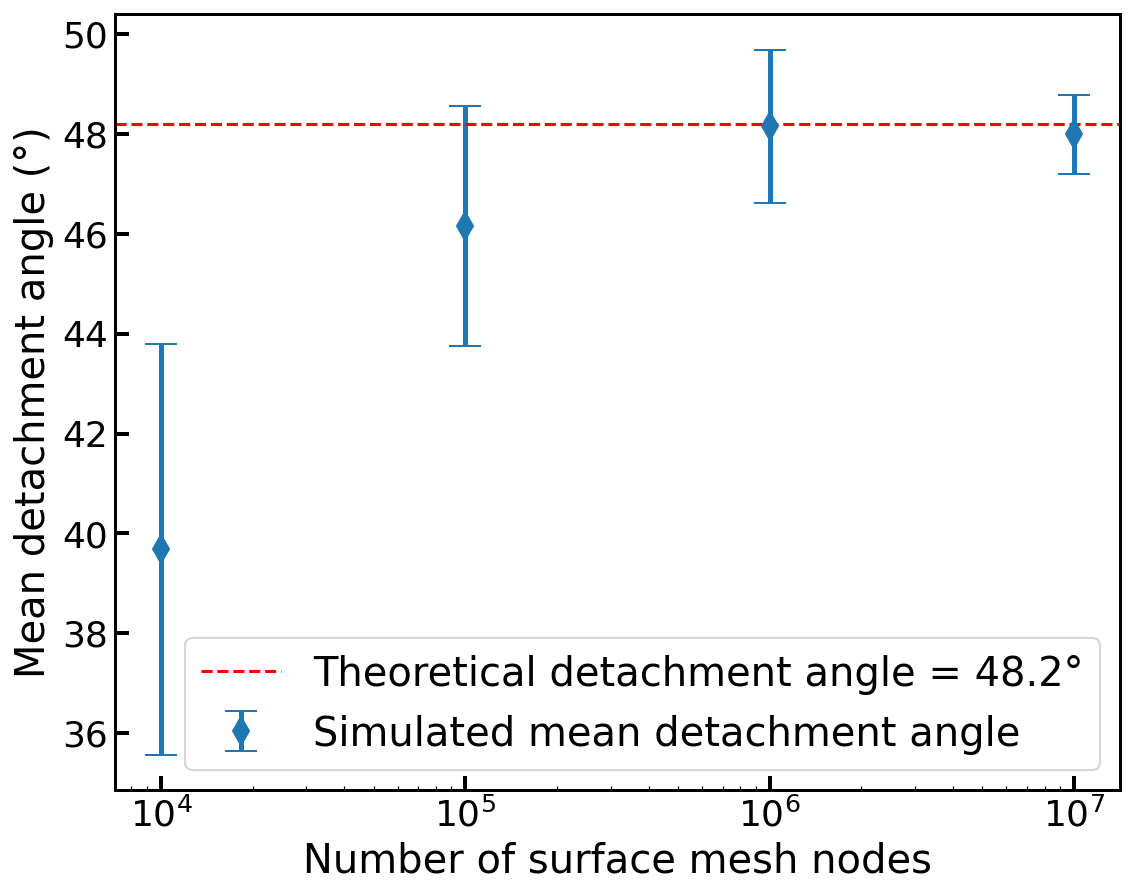}

  \caption{Detachment angle of particles sliding without friction on a polyhedral sphere under uniform gravity, as a function of mesh resolution. The theoretical value for a smooth sphere is 48.189$^\circ$ (red dashed line). As the mesh resolution increases, the mean measured detachment angle computed by RAVEL converges towards the analytical prediction, while the dispersion decreases. Error bars indicate one standard deviation.}

  \label{Figure_annex_test_convergence}

\end{figure}

 {We repeated the experiment with meshes containing $10^4$, $10^5$, $10^6$, and $10^7$ nodes (Figure \ref{Figure_annex_test_convergence}).  }

 {The mean measured values of the detachment angle are 39.68$^\circ$, 46.16$^\circ$, 48.15$^\circ$ and 47.99$^\circ$ respectively, with standard deviations decreasing with increasing number of nodes: 4.12$^\circ$, 2.4$^\circ$, 1.54$^\circ$ and 0.793$^\circ$. At the highest resolution, the numerical result agrees with the analytical prediction within 0.20$^\circ$. Note that, since our particles are released  within $\theta_0=$ 1$^\circ$ of the sphere's North Pole,  the theoretical detachment angle is slightly modified since $\alpha=arccos\big[(2/3)cos~\theta_0\big]$ (derived from \citep{Prior2007_sliding_sphere}) giving $\alpha=48.197 ^\circ$ instead of $\alpha=48.189 ^\circ$.
This test confirms that RAVEL accurately reproduces particle dynamics on curved surfaces when the mesh is sufficiently resolved.}

\section{Visual evidence of topographically controlled ejecta deposition}
\label{annexe_topographic_effects}

 {In both the DART/DRACO-based and synthetic fully rough DTMs, the final ejecta distributions exhibit pronounced spatial heterogeneities, characterized by elongated concentrations of deposited material (streaks) and extended regions depleted of particles (see, e.g., Fig.~\ref{fig:3D_dart_model}). Here, we investigate the role of local surface topography in the formation of these structures by examining individual ejecta trajectories and their interactions with the terrain using the synthetic fully rough Dimorphos DTM. This model was specifically designed to incorporate topographic features across the entire surface, making it particularly well suited to illustrating how local topography influences particle trajectories and surface transport on Dimorphos.}

\begin{figure*}

  \centering

  \includegraphics[width=0.9\linewidth]{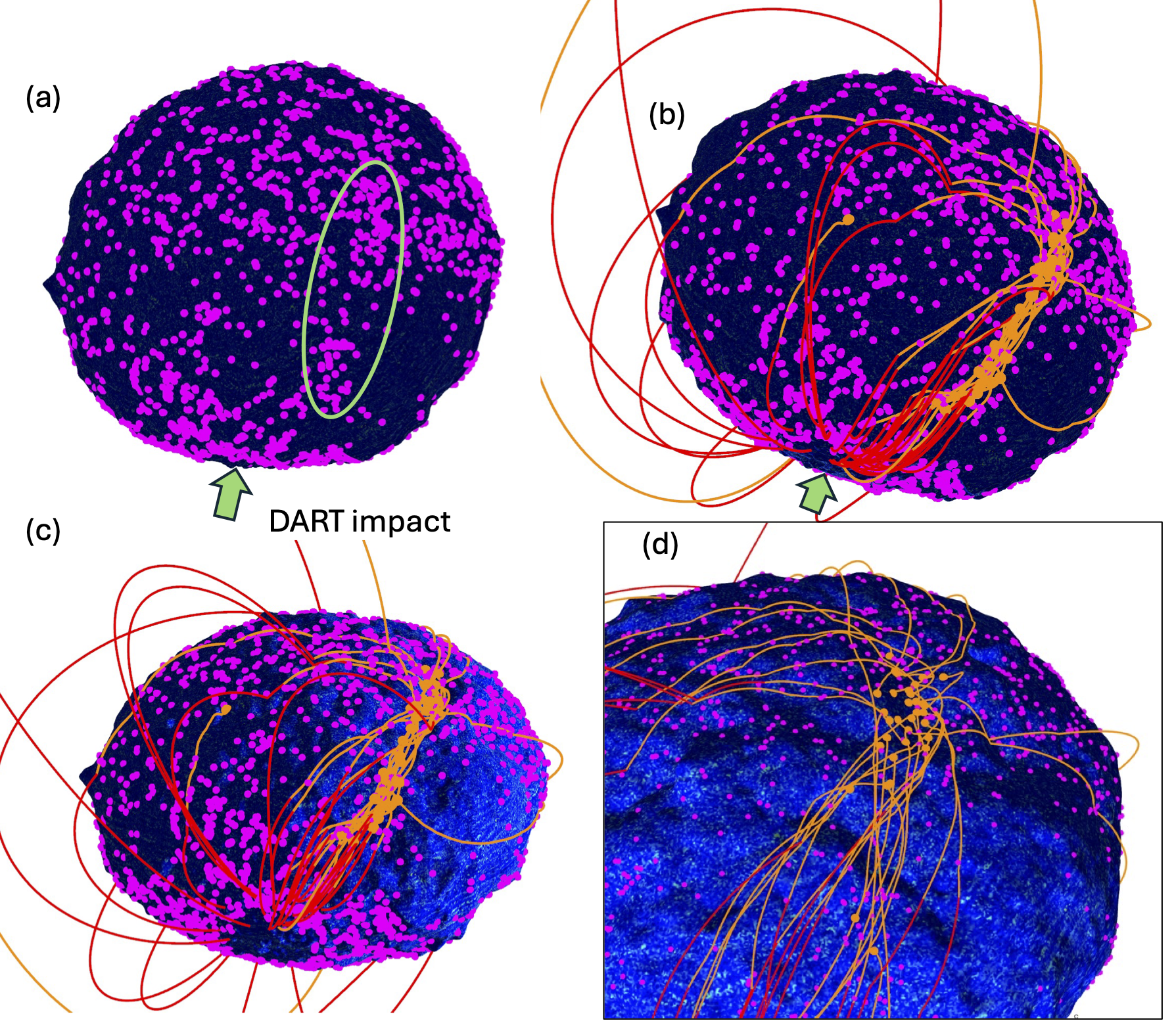}

  \caption{ Formation of an ejecta streak following the DART impact, using a synthetic fully rough shape model of Dimorphos. The surface is coloured according to the local dynamic slope, with dark blue corresponding to low slopes. Magenta dots indicate the final resting positions of ejecta particles. Selected particle trajectories are shown in red during their initial ballistic phase, before their first surface contact, and in orange after their first contact, including subsequent ballistic excursions and surface motion. The green arrow indicates the DART impact location. Dimorphos is viewed from its southern hemisphere. Different illumination geometries are used to highlight the local topography. (a) Global view of the final ejecta distribution. The green ellipse highlights an elongated streak of deposited particles. The surface is displayed with reduced illumination to facilitate the identification of the magenta points; (b) Trajectories of selected particles contributing to the streak. Several particles undergo extended ballistic motion, sometimes completing nearly one orbit around Dimorphos before reaching the surface. After their first contact, their trajectories exhibit a preferential alignment towards the streak, suggesting a channelling effect; (c) Same as panel (b), but with grazing illumination to emphasize the surface relief. The streak appears to follow an elongated topographic depression resembling a shallow canyon or gorge; (d) Close-up of the streak region. The post-impact trajectories show coherent changes in direction associated with the local relief, with several particles following preferential paths through valley-like topographic structures.}

\label{Figure_zoom_on_streak}

\end{figure*}

\begin{figure*}

  \centering

  \includegraphics[width=0.9\linewidth]{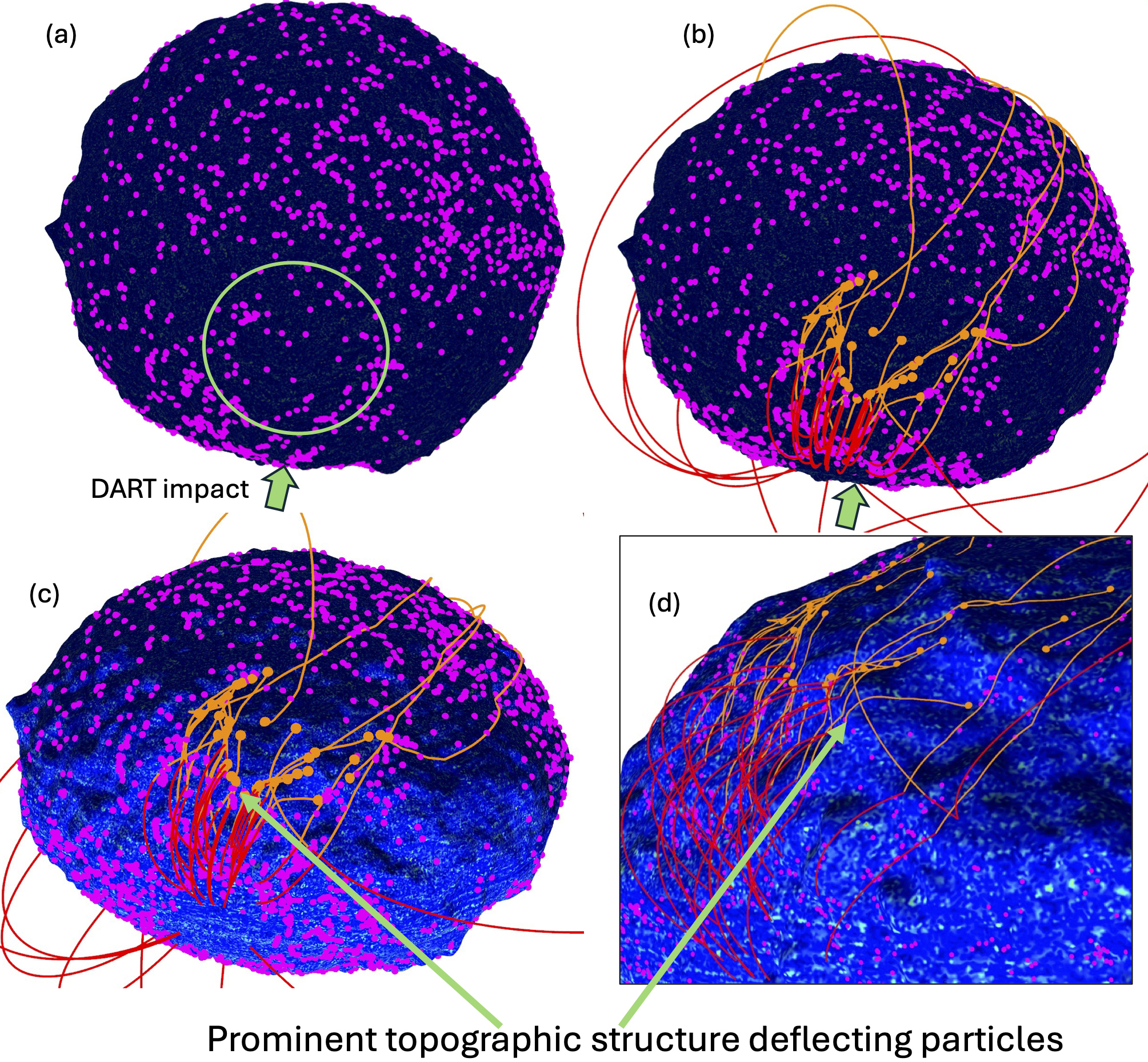}

 \caption{ Formation of a topographically shadowed region in the ejecta deposition
pattern following the DART impact, using a synthetic fully rough shape
model of Dimorphos. The colour conventions are the same as in
Fig.~\ref{Figure_zoom_on_streak}: magenta dots indicate the final
resting positions of ejecta particles, while red and orange lines
represent particle trajectories before and after their first surface
contact, respectively. The surface is coloured according to the local dynamic
slope, and the short green arrow in panel (a) indicates the DART impact
location. Dimorphos is viewed from its southern hemisphere.
(a) Global view of the ejecta deposition pattern.
The green circle highlights a region characterized by a pronounced
depletion of deposited particles; 
(b) Trajectories of selected particles in the vicinity of
the depleted region. Several trajectories exhibit significant
deflections following their first surface contact; 
(c) Same region viewed under grazing illumination to
emphasize the local topography. The trajectories reveal that
several particles interact with a prominent topographic feature
located near the boundary of the depleted region.
Following these interactions, particles are deflected and may
travel ballistically above the lower-lying terrain behind the obstacle,
reducing the likelihood of deposition in this region.
The long green arrow identifies the topographic feature responsible
for these deflections;
(d) Close-up of the prominent topographic feature
indicated in panel (c). Several particle trajectories exhibit
coherent changes in direction in its vicinity, illustrating
how local relief can redirect ejecta and produce a
topographic shadow in the final deposition pattern.}

\label{Figure_zoom_on_shadowed_region}

\end{figure*}

 {We illustrate here that the spatial patterns of ejecta deposition arise from the dynamical interplay between particle trajectories and the irregular surface topography of Dimorphos.
Figure~\ref{Figure_zoom_on_streak} shows how successive surface
interactions redirect particles towards preferential pathways,
producing elongated concentrations of deposited ejecta along valley-like structures.
Conversely, Fig.~\ref{Figure_zoom_on_shadowed_region} illustrates
how prominent topographic features deflect particles away from
adjacent regions, producing depleted areas.
The coherent deflection of independent particle trajectories near
the same topographic features, together with the absence of such
structures on a smooth ellipsoid, strongly supports their physical
origin rather than a purely numerical artefact.}

\end{document}